\documentclass[lettersize,journal]{IEEEtran}

\usepackage{bm}
\usepackage{amsmath,amssymb,amsfonts}
\usepackage{bbm}
\usepackage{mathtools}
\usepackage{upgreek}
\usepackage{cases}
\usepackage{xintexpr}
\usepackage{pifont}
\usepackage{ntheorem}
\usepackage{booktabs}
\usepackage{multirow}
\usepackage{graphicx}
\usepackage{subcaption}
\usepackage{overpic}
\usepackage{textcomp}
\usepackage{xcolor}
\usepackage{xifthen}
\usepackage{multicol}
\usepackage{cuted}
\usepackage{cite}
\usepackage{enumitem}
\usepackage[switch]{lineno}
\usepackage[numbers,sort&compress]{natbib}
\usepackage[ruled,linesnumbered,noend]{algorithm2e}
\usepackage{hyperref,soul}
\usepackage{cancel}
\usepackage{nicematrix}
\DeclareMathAlphabet{\mathpzc}{OT1}{pzc}{m}{it}

\def\BibTeX{{\rm B\kern-.05em{\sc i\kern-.025em b}\kern-.08em
    N\kern-.1667em\lower.7ex\hbox{E}\kern-.125emX}}

\newcommand{\highlight}[1]{\vspace{1mm}\noindent{}\textbf{#1}}

\newcommand{\w}[1][]{
\ifthenelse{\isempty{#1}}
{\mathbf{w}}
{\mathbf{w}^{(#1)}}}
\newcommand{\tw}[1][]{
\ifthenelse{\isempty{#1}}
{\widetilde{\mathbf{w}}}
{\widetilde{\mathbf{w}}^{(#1)}}}
\newcommand{\hw}[1][]{
\ifthenelse{\isempty{#1}}
{\widehat{\mathbf{w}}}
{\widehat{\mathbf{w}}^{(#1)}}}

\newcommand{\vp}{\mathbf{p}}
\newcommand{\x}{\mathbf{x}}
\newcommand{\y}{\mathbf{y}}

\newcommand{\set}{\mathcal{S}}

\newcommand{\update}[1][k]{\Delta^{(#1)}}

\newcommand{\hupdate}[1][k]{\hat{\Delta}^{(#1)}}

\newcommand{\vm}[1][k]{\mathbf{m}^{(#1)}}
\newcommand{\tvm}[1][k]{\tilde{\mathbf{m}}^{(#1)}}
\newcommand{\bvm}[1][k]{\bar{\mathbf{m}}^{(#1)}}
\newcommand{\hvm}[1][k]{\hat{\mathbf{m}}^{(#1)}}
\newcommand{\vrho}[1][k]{\boldsymbol{\rho}^{(#1)}}

\newcommand{\mG}{\mathbf{G}}

\newcommand{\argmin}{\mathop{\mathrm{argmin}}\limits} 

\newcommand{\expect}[1][]{
\ifthenelse{\isempty{#1}}
{\mathbb{E}}
{\mathbb{E}\left[#1\right]}}

\newcommand{\fw}[1][]{
\ifthenelse{\isempty{#1}}
{\mathbf{\bar{w}}}
{\mathbf{\bar{w}}^{(#1)}}
}

\newcommand{\pdist}[1][m]{\mathcal{P}_{m}}

\newcommand{\p}[1][]{
\ifthenelse{\isempty{#1}}
{\boldsymbol{p}}
{\boldsymbol{p}^{(#1)}}
}

\newcommand{\intab}[2][0.75]{
\scalebox{#1}{\textrm{#2}}
}

\newcommand\addpicture[3]{%
\setbox\mybox=\hbox{\includegraphics[scale=#3]{#2}}
\myboxwidth\wd\mybox    
\renewcommand\windowpagestuff{%
\includegraphics[scale=#3]{#2}
\captionof{figure}{A test figure.}}
\parpic[#1]{%
\begin{minipage}{\myboxwidth}
 \windowpagestuff 
\end{minipage} 
} }

\usepackage{tikz}

\usepackage{tcolorbox}

\definecolor{fedvote}{RGB}{235,255,251}
\definecolor{byzantinefedvote}{RGB}{235,246,255}

\ifodd 1

\newcommand{\com}[1]{\textbf{\color{blue}([KY]: #1)}}
\newcommand{\comRJ}[1]{\textbf{\color{red}([RJ]: #1)}}
\newcommand{\CommentWong}[1]{\textcolor[rgb]{1,0,0}{[Wong: #1]}}
\newcommand{\comHD}[1]{\textbf{\color{red}([HD]: #1)}}
\else

\newcommand{\com}[1]{}
\newcommand{\comRJ}[1]{}
\newcommand{\CommentWong}[1]{}
\newcommand{\comHD}[1]{}
\fi

\begin{document}
\title{Advancing Hybrid Defense for Byzantine Attacks in Federated Learning}
\author{Kai Yue, \textit{Graduate Student Member, IEEE}, Richeng Jin, \textit{Member, IEEE}, Chau-Wai Wong, \textit{Senior Member, IEEE}, and Huaiyu Dai, \textit{Fellow, IEEE}}


\maketitle
\makeatletter{\renewcommand*{\@makefnmark}{}


\makeatother}

\begin{abstract}

Federated learning (FL) enables multiple clients to collaboratively train a global model without sharing their local data. 
Recent studies have highlighted the vulnerability of FL to Byzantine attacks,  where malicious clients send poisoned updates to degrade model performance. 
In particular, many attacks have been developed targeting specific aggregation rules, whereas various defense mechanisms have been designed for dedicated threat models. 
This paper studies the resilience of attack-agnostic FL scenarios, where the server lacks prior knowledge of both the attackers' strategies and the number of malicious clients involved. 
We first introduce hybrid defenses against state-of-the-art attacks.
Our goal is to identify a general-purpose aggregation rule that performs well on average while also avoiding worst-case vulnerabilities.
By adaptively selecting from available defenses, we demonstrate that the server remains robust even when confronted with a substantial proportion of poisoned updates.
We also emphasize that existing FL defenses should not automatically be regarded as secure, as demonstrated by the newly proposed Trapsetter attack.
The proposed attack outperforms other state-of-the-art attacks by further increasing the impact of the attack by $5$--$15\%$. 
Our findings highlight the ongoing need for the development of Byzantine-resilient aggregation algorithms in FL.  

\end{abstract}

    
\section{Introduction}

Federated learning (FL) is a privacy-preserving framework in which multiple clients jointly optimize a machine learning model~\cite{kairouz2021advances}.
Under this framework, a central server coordinates the training process by aggregating individual clients' local models, which are updated based on their private data.
Many efforts have been devoted to the design of server aggregation algorithms. 
Federated averaging~(FedAvg) is a popular scheme in which clients update their local models using algorithms such as stochastic gradient descent~(SGD)~\cite{mcmahan2017communication}.
The server then updates the global model by computing the average of the model weights. 
Despite advances in aggregation algorithms to handle unbalanced and non-independent and identically distributed~(non-IID) data~\cite{karimireddy2020scaffold, zhao2018federated}, FL faces many other significant challenges, particularly vulnerability to Byzantine attacks~\cite{blanchard2017machine}.
This class of threats involves clients acting arbitrarily or maliciously, in order to disrupt model convergence~\cite{lyu2022privacy, shi2022challenges, shejwalkar2022back}.
FedAvg and its variants are known to be vulnerable to Byzantine attacks, wherein a small fraction of adversaries can significantly degrade model performance~\cite{blanchard2017machine, yin2018byzantine}.
Although substantial research efforts have been dedicated to FL resilience, the ongoing cat-and-mouse game between attackers and defenders has only intensified.

In Byzantine attacks, malicious clients can manipulate model updates by using poisoned data or sending poisoned gradients to the server.
Specifically, attackers can poison data labels~\cite{fang2020local, karimireddy2021learning}, inject noise into gradients and mislead the server~\cite{xie2020fall, li2019rsa}, and build optimization processes to hide poisoned updates~\cite{shejwalkar2021manipulating}. 
These attacks often assume strong knowledge of benign clients' updates or the server's aggregation algorithm.
Therefore, some recent studies suggest that the severity of existing Byzantine attacks may have been overstated~\cite{shejwalkar2022back}.

In response to Byzantine attacks, defenders have proposed various aggregation rules based on robust statistics to mitigate their impact, including the median aggregation, trimmed mean~\cite{yin2018byzantine}, and Krum~\cite{blanchard2017machine}. 
More advanced methods exploit detection and filtering techniques, such as sign-guided majority vote~\cite{bernstein2018signsgd, xu2022byzantine}, outlier detection and rejection~\cite{sattler2020byzantine, shejwalkar2021manipulating}, and personalized training~\cite{sattler2020byzantine, yu2020salvaging}. 
However, these aggregation rules are typically tailored for specific threat models or based on oversimplified setups that do not fully account for the coexistence of data heterogeneity in federated learning~\cite{li2024blades}.
Although many works have claimed that the defender does not have knowledge about the attack, their proposed defense turns out to implicitly rely on restricted threat models~\cite{fang2024byzantine, yan2025skymask}. 
Meanwhile, the theoretical convergence guarantees of existing defenses may not translate into real-world performance~\cite{shejwalkar2022back}. 
Consequently, understanding the impact of attackers in FL is still an open problem.

Essentially, FL resilience depends on the ongoing arms race between attackers and defenders. 
In real-world scenarios, attackers may adjust their tactics if they realize that current ones are ineffective. 
Furthermore, the proportions of attackers and benign clients can fluctuate over time.
To address this challenge, defenders may remain agile and use adaptive aggregation rules to ensure the stability of FL systems. 

In this paper, we evaluate various Byzantine attacks and defenses in an attack-agnostic setting, where the server has little knowledge of attackers' tactics or proportions.
We focus on identifying a general-purpose aggregation rule that remains effective across diverse scenarios while avoiding worst-case vulnerabilities. 
To this end, we propose a hybrid defense that adaptively chooses aggregation rules to achieve Byzantine resilience.
We also propose the Trapsetter attack strategy that dynamically navigates the model optimization landscape and misleads the server toward a solution that has a relatively small distance from the original one, but results in poor performance. Our findings underscore the importance of continually evolving Byzantine-resilient algorithms for FL systems. 
The contributions are summarized as follows:

\begin{enumerate}
    \item[\textbullet]
    To the best of our knowledge, we are among the first to comprehensively investigate attack-agnostic scenarios in FL. 
    Our case studies demonstrate that none of the state-of-the-art approaches can guarantee system robustness and each defense has its own Achilles' heel. 

    \item[\textbullet] 
    The proposed hybrid defense enhances Byzantine resilience in attack-agnostic FL settings. This not only improves the model's average performance, but also mitigates the impact of worst-case vulnerabilities.

    \item[\textbullet] 
    The proposed Trapsetter attack strategy shows its effectiveness across different tasks under the stronger hybrid defense scheme.

\end{enumerate}

The remainder of the paper is organized as follows. Section~\ref{section:preliminaries} reviews the background on FL, Byzantine attacks, and robust defenses. Section~\ref{section:defense} presents our proposed hybrid defense framework and evaluates its performance under various attack-agnostic scenarios. In Section~\ref{section:attack}, we introduce the TrapSetter attack and demonstrate its ability to challenge the robustness of FL. Section~\ref{section:conclusion} concludes the paper with a summary of the findings and future research directions.

\section{Preliminaries}\label{section:preliminaries}
In this section, we will review the details of FL, Byzantine attacks, and defense schemes.
The symbol conventions used are as follows. 
We use $[N]$ to denote a set of integers $\{1, 2,\dots,N\}$. 
Lowercase boldface letters, such as $\x$ and $\w$, are used to denote column vectors, while calligraphic letters, such as $\mathcal{A}$ and $\mathcal{M}$, are used to denote sets. 
TABLE~\ref{tab:notation} summarizes the key notations used in this paper.

\begin{table}
    \centering
    \caption{Notations}\label{tab:notation}
    \begin{tabular}{ll}
        \toprule
        \textbf{Notation}  &  \textbf{Description}  \\
        \midrule 
        $A$ & number of attackers \\
        $B$ & number of benign clients \\
        $\mathcal{A}$ & attacker set \\
        $\mathcal{B}$ & benign client set \\
        $a$ & attacker index \\
        $b$ & benign client index \\
        $d$ & model weight dimension \\
        $\eta$ & local update learning rate \\
        $k$ & communication round index \\
        $m$ & general client index (benign/malicious) \\
        $\vm_b$ & local momentum for client $b$ at round $k$\\
        $M$ & total number of clients \\
        $N_m$ & number of training examples for client $m$ \\
        $\vp^{(k)}$ & perturbation vector at round $k$ \\
        $t$ & local optimization step index \\
        $\tau$ & number of local update steps \\
        $(\x_{m,i}, y_{m,i})$ & $i$th  data  pair for client $m$ \\ 
        $\w$ & vectorized model weights \\
        $R(\w;\x,\y)$ & sample-wise empirical risk function \\
        $F_{m}(\w)$ & local objective function for client $m$ \\
        $\update[k,t]_m$ & local update at round $k$ and step $t$ on client $m$ \\
        $\update[k]$ & global model update of FedAvg at round $k$ \\
        \bottomrule
        \end{tabular}
    \end{table}

\subsection{Federated Learning}
In FL, a machine learning model is trained in multiple decentralized entities, each holding local data samples.
Following~\cite{mcmahan2017communication, kairouz2021advances}, we consider an FL architecture in which a server optimizes a model by coordinating $M$ clients. 
The local dataset of the $m$th worker is denoted by $\mathcal{D}_{m} \triangleq \{(\x_{m,i}, y_{m,i})\}_{i=1}^{N_m}$, where $(\x_{m,i}, y_{m,i})$ represents an input--output pair and $N_m$ is the number of training examples for the worker $m$. 
The goal of each worker is to minimize the local objective, formulated as empirical risk minimization in training examples $N_m$,

\begin{equation}
F_{m}(\w) \triangleq \frac{1}{N_m} \sum_{i=1}^{N_m} R(\w; \x_{m,i}, y_{m,i}), 
\end{equation}
where $R(\cdot;\cdot)$ is a sample-wise risk function quantifying the model's error, with $\w \in \mathbb{R}^{d}$ being the model's weight vector. 
For simplicity, the risk function on dataset $\mathcal{D}_m$ is also abbreviated as $R(\w; \mathcal{D}_m)$.
The global objective function, $F(\w)$, aims to find the optimal weights $\w$ that minimize the averaged loss across all clients:
\begin{equation}\label{eq:global_obj}
F(\w) \triangleq \frac{1}{M} \sum_{m=1}^{M} F_{m}(\w),
\end{equation}
where $M$ is the total number of clients. 
To solve the optimization problem in \eqref{eq:global_obj}, FedAvg~\cite{mcmahan2017communication} aggregates model updates from selected clients and updates a global model. 
In each communication round $k$, the server broadcasts the current global model weights $\w[k]$ to clients. 
Upon receiving the global model, each client $m$ initializes the local model with broadcasted weights, denoted by $\w[k,0]_m$, and begins to optimize it via multiple steps of stochastic gradient descent (SGD).
Denote a mini-batch of size $n_m$ as $\xi^{(k,t)}_{m}$, sampled uniformly randomly from $\mathcal{D}_m$, where $t$ is the local step index.
The local weight at step $\tau$ may be formulated as:
\begin{equation}
    \w[k,\tau]_{m}\! = \w[k,0]_m \!-\! \frac{\eta}{N_m} \sum_{t=0}^{\tau-1}\! \nabla R (\w[k,t]_m; \xi^{(k,t)}_{m} ), 
\end{equation}
where $\eta$ is the learning rate. 
In this context, the term ``gradient'' or ``update'' is also used to refer to the model difference $\update[k, \tau]_m \triangleq \w[k,0]_m - \w[k,\tau]_m$, representing the total change in the model's weights after $\tau$ steps of local optimization. 
This cumulative update is analogous to a gradient in the sense that it guides the update of the global model.
The global update via FedAvg is computed as:
\begin{equation}
    \update[k] = \frac{1}{M} \sum_{m=1}^{M} \update[k, \tau]_m.
\end{equation}
The global model is  updated via $\w[k+1] = \w[k] - \eta_{g}\update[k]$, where $\eta_{g}$ is the global learning rate.

\subsection{Poisoning Attacks}
We begin by describing the threat model, which covers the objectives, knowledge, and capabilities of the adversaries. 
Subsequently, we review several state-of-the-art poisoning attacks proposed in the literature.
For clarity, we use the index $a \in \mathcal{A}$ to denote attackers, where $\mathcal{A}$ represents the attacker set. 
Similarly, we use $b \in \mathcal{B}$ to denote benign clients, with $\mathcal{B}$ representing the set of benign clients.
The number of attackers is denoted as $A \triangleq |\mathcal{A}|$, and the number of benign clients is $B \triangleq |\mathcal{B}|$. 
For simplicity, we assume that all attackers behave the same unless otherwise specified. 
The attacker datasets $\mathcal{D}_{a}$, $a \in \mathcal{A}$, can be identical or follow the same distribution, which will be specified in the simulation.

\highlight{Threat Model. } 
We consider client-side attackers capable of arbitrarily manipulating their local data or updates. 
The objective of these attackers is to compromise the integrity of the global model by skewing the optimization process without being detected. 
There are two widely recognized attack types based on a fine-grained objective taxonomy~\cite{shejwalkar2022back}.
The first type is \textit{targeted attack}, which induces the model to misclassify certain inputs to target labels specified by the adversaries. 
The second type is \textit{untargeted attack}, which aims to reduce the overall classification accuracy across all labels. 
In terms of adversarial knowledge, we assume a worst-case scenario in which attackers have full access to all client data. 
Furthermore, as participants in the FL system, attackers are assumed to have complete knowledge of the architecture and parameters of the global model.

In this context, a defender is assumed to be the FL server, which aggregates updates received from participating clients. 
In \textbf{attack-agnostic settings}, the server does not have prior knowledge of the distribution of attackers or their tactics. 
In particular, attackers may form different groups and adopt different attack algorithms.
They may send totally different poisoned messages within the same communication round, leaving the defender unprepared.  
For example, in one communication round, attackers may form distinct groups, with one group launching a targeted attack while another executes an untargeted attack. 
An aggregation scheme that is robust against one attack group may fail to defend against the other.
Additionally, attackers may collude by uploading decoy models to mislead the server, enabling other attacker groups to persist with their malicious updates. 
Attackers may also vary their strategies over time, switching attack algorithms between different communication rounds to further evade detection. 
This dynamic behavior significantly increases the complexity of resilient FL in real-world applications. 
Practical attack scenarios are elaborated in Section~\ref{sec:hybrid_defense}. 
We now review several state-of-the-art attack algorithms in the following to provide more context for the types of adversarial strategies that defenses must address.

\highlight{Inner Product Manipulation (IPM)~\cite{xie2020fall}.}
IPM is an untargeted attack. 
IPM attackers disrupt the learning process by achieving negative inner products between the mean of benign updates and aggregated updates. 
For each IPM attacker, the poisoned update can be designed as: 
\begin{equation}\label{eq:ipm}
\Delta^{(k,\tau)}_{a} = -\frac{\epsilon}{B} \sum_{b \in \mathcal{B}} \update[k, \tau]_b , \quad a \in \mathcal{A},
\end{equation}
where $\epsilon$ is a positive coefficient that dictates the scale of the malicious updates. 
When $\epsilon > \frac{B}{A}$, the aggregated update based on FedAvg is  
\begin{equation}
    \update[k] = \frac{B - A\epsilon}{MB} \sum_{b \in \mathcal{B}} \update[k, \tau]_b. 
\end{equation}
We note that $B - A\epsilon < 0$, which results in a negative inner product between the mean of benign updates and aggregated updates,
\begin{equation}
    \left\langle \update[k],  \frac{1}{B}\sum_{b \in \mathcal{B}}\update[k, \tau]_b  \right\rangle \leqslant 0.
\end{equation}

\highlight{Minimize Maximum Distance (Min-Max)~\cite{shejwalkar2021manipulating}.}
MinMax is an untargeted attack. 
MinMax attackers add a perturbation vector $\vp^{(k)}$ to the mean of benign gradients,
\begin{equation}
    \update[k,\tau]_a = \frac{1}{B} \sum_{b \in \mathcal{B}} \update[k, \tau]_b + \gamma \, \vp^{(k)}. 
\end{equation}
The scalar $\gamma$ can be obtained by solving an optimization problem 
\begin{subequations}
\begin{align}
    \underset{\gamma}{\operatorname{argmax}} \;&\; d(\update[k,\tau]_a, \update[k,\tau]_{\mathcal{B}}) \triangleq \max_{b \in \mathcal{B}}\left\|\update[k,\tau]_a - \update[k,\tau]_b\right\|_2, \\
    \text{s.t.}  \;&\;  d(\update[k,\tau]_a, \update[k,\tau]_{\mathcal{B}}) \leqslant \max _{b_1, b_2 \in \mathcal{B}}\left\|\update[k,\tau]_{b_1} - \update[k,\tau]_{b_2}\right\|_2. 
\end{align}
\end{subequations}
This optimization ensures that the maximum distance between any malicious update and benign updates does not exceed the largest distance observed among benign updates. 
As a result, attackers can dynamically adjust the noise, while reducing the risk of being detected. 

\highlight{Relocated Orthogonal Perturbation (ROP)~\cite{ozfatura2023byzantines}.}
ROP is an untargeted attack that leverages the knowledge of local momentum. 
A local momentum term for a benign client $b$ at communication round $k$ is represented as:
\begin{equation}
    \vm_b = (1-\beta) \update[k, \tau]_b + \beta \vm[k-1]_{b},
\end{equation}
where the iteration starts with a zero vector, $\vm[0]_{b} = \mathbf{0}$, and $\beta \in [0,1)$ is a scaling factor. 
When $\beta$ is $0$, the momentum term is equivalent to the local update $\update[k, \tau]_b$.
ROP attackers utilize the aggregated momentum term  from benign clients, 
\begin{equation}
    \tvm[k] \triangleq \frac{1}{B} \sum_{b \in \mathcal{B}} \vm_b.
\end{equation}
Suppose that the aggregated momentum across all clients is
$    \bvm[k] \triangleq \frac{1}{M} \sum_{i=1}^M \vm_i.$
ROP first chooses a reference vector between $\tvm[k-1]$ and $\bvm[k]$, 
\begin{equation}
    \hvm[k] = \lambda \tvm[k-1] + (1-\lambda)\bvm[k],
\end{equation}
where $\lambda \in (0,1)$ is a scaling factor. 
The attackers then generate a vector $\vrho$ orthogonal to $\hvm[k]$. 
The poisoned update is given by 
\begin{equation}
    \update[k]_a = \sin (\Pi) \frac{\vrho}{\| \vrho \|} + \cos (\Pi)\frac{ \hvm }{\| \hvm \|} ,
\end{equation}
where $\Pi \in (0, 2\pi)$ is the angle between the reference and the poisoned update.
This method avoids direct opposition to the reference direction $\hvm[k]$. 
By tuning $\Pi$, the attackers take advantage of momentum and add a perturbation that is less detectable and potentially more disruptive over time.

\highlight{Sign Flipping (SF)~\cite{li2019rsa}.}
SF is an untargeted attack. 
Contrary to the aforementioned attacks, SF attackers do not need to access benign clients' gradients. 
Instead, they increase the loss via gradient ascent, flipping the signs of the local update.
Attackers may scale up the update to increase the impact.

\highlight{3DFed~\cite{li20233dfed}.}
3DFed is a targeted poisoning attack composed of three evasion modules. 
In the first module, adversaries conduct backdoor training with constrained optimization, specifically designed to evade defenses that rely on gradient norm thresholds. 
The second module introduces optimized noise masks that conceal backdoor features within the poisoned model. 
The third module facilitates collusion among attackers by generating decoy models, effectively circumventing dimension-reduction-based defenses.
Each module employs iterative optimization to ensure that the attack remains adaptive during each communication round. 
To avoid being detected easily, 3DFed identifies redundant neurons within the neural network and uses timely feedback to adapt the hyperparameters in the attack algorithm.

\highlight{Neurotoxin (NT)~\cite{zhang2022neurotoxin}.}
Similar to 3DFed, NT is a targeted attack that leverages poisoned datasets and poisoned gradients to compromise the model. 
NT attackers first obtain benign updates and compute gradients using poisoned datasets. 
They then identify the bottom$\omega\%$ coordinates of benign updates based on their magnitudes and apply the projected gradient descent to these coordinates.
These coordinates are treated as redundant neurons, effectively concealing the poisoning noise within the model.

\subsection{Robust Defense}\label{section:defense}
Next, we review state-of-the-art defense mechanisms proposed in the literature to mitigate the impact of poisoning attacks.
The input to the defender consists of a set of updates that may include both benign and malicious contributions. 
With a slight abuse of notation, we use $\update[k]_{\mathcal{C}} \triangleq \{ \update[k, \tau]_{a_1}, \dots, \update[k, \tau]_{a_{A}}, \update[k, \tau]_{b_{1}}, \dots,  \update[k, \tau]_{b_{B}}\}$ to denote the set of updates received in the $k$th round. 
Formally, the defender aims to design an aggregation function $f(\cdot ; \cdot)$ that gives a robust update $\hupdate[k] \in \mathbb{R}^{d}$, i.e.,
\begin{equation}\label{eq:general_f}
    \hupdate[k] = f(\update[k]_{\mathcal{C}}; \mathcal{I}), 
\end{equation}
where $\mathcal{I}$ represents additional information available to the defender, which may include attack-specific details or a trusted reference. 
Based on the assumptions of defenders' knowledge in existing works, we categorize them into two types.
The first type is \textit{informed} defender, which knows the attack information, for example, the number/fraction of attackers, attacker tactics, or hyperparameters in the attack algorithm. 
The second type is \textit{uninformed} defender, which does not explicitly require additional knowledge or assumptions about the attack.

Although the terms ``uninformed'' and ``attack-agnostic'' defenders may appear similar, it is important to distinguish the concept of an ``uninformed'' defender from the ``attack-agnostic'' defender investigated in this work. 
Many existing uninformed defenders may implicitly rely on assumptions about the attacker's tactics or add constraints on their behavior based on the specific mechanisms implemented. 
As a result, they may perform poorly in attack-agnostic settings and cannot be regarded as effective attack-agnostic defenses.
For example, norm-based criteria employed by the defenders~\cite{blanchard2017machine, karimireddy2021learning, fang2024byzantine}, regardless of whether the defender is informed or not, inherently assume that the attackers alter the magnitude of gradients. 
Similarly, frequency-based criteria~\cite{fereidooni2024freqfed}, though seemingly uninformed, assume that attackers concentrate their manipulations on high-frequency components. 
These defenses may falter against any attack that deviates from implicit assumptions. 
For defenders that do not require additional input, we simplify~\eqref{eq:general_f} by setting $\mathcal{I} = \emptyset$ and omitting the second input argument. 
The following sections provide a review of well-received defenses in the literature.

\highlight{Balance~\cite{fang2024byzantine}.}
Balance is an uninformed defense that relies on an evaluation dataset for reference. 
Suppose that the defender calculates a reference update $\update_{\text{ref}}$ based on the dataset $\mathcal{D}_{\text{ref}}$.
For each received gradient $\update[k,\tau]_{m}$, the defender chooses to accept it if the following inequality holds, 
\begin{equation}\label{eq:balance}
    \| \update[k]_{\text{ref}} - \update[k,\tau]_{m} \| \leqslant \phi \exp\left[- \kappa \lambda(k) \right] \| \update[k]_{\text{ref}} \|,  
\end{equation}
where $\phi>0$ is a scaling factor, $\kappa>0$ controls how fast the exponential function decays, and $\lambda(k) = k/K$ is a monotonic increasing function with respect to communication round index $k$. 
After a benign set $\mathcal{S}_{\text{B}}$ is selected based on $\eqref{eq:balance}$, the aggregation results can be obtained via FedAvg, 
\begin{equation}
    f_{\text{Balance}}(\update[k]_{\mathcal{C}}; \mathcal{D}_{\text{ref}}) = \frac{1}{|\mathcal{S}_{\text{B}}|}\sum_{b \in \mathcal{S}_{\text{B}}} \update[k,\tau]_{b}. 
\end{equation}

\highlight{Centered Clipping (CC)~\cite{karimireddy2021learning}.}
CC is an uninformed defense.
CC defender leverages momentum $\bvm$ throughout the communication rounds to scale the current updates received. 
To better understand how it works, consider the following centered clipping operation
\begin{equation}
    g_{\text{CC}}(\mathbf{m}; \bar{\mathbf{m}}, \rho) = \bar{\mathbf{m}} + \min \left(1, \frac{\rho}{\|\mathbf{m} - \bar{\mathbf{m}}\|}\right) (\mathbf{m}-\bar{\mathbf{m}}),
\end{equation}
where $\mathbf{m}$ is an input vector, $\bar{\mathbf{m}}$ is a reference vector, and the radius $\rho$ is a positive scaling parameter. 
In general, if the input $\mathbf{m}$ is close to the reference, i.e., $\|\mathbf{m} - \bar{\mathbf{m}}\| < \rho$, $f_{\text{CC}}$ recovers the identity function. 
Otherwise, the input is scaled to the reference.
The defender applies $f_{\text{CC}}$ to each received update and then averages the result: 
\begin{equation}
    f_\text{CC}(\update[k]_{\mathcal{C}}) = \frac{1}{M} \sum_{m=1}^{M} g_{\text{CC}}(\update[k]_m; \bvm[k], \rho).
\end{equation}
Regarding the hyperparameter $\rho$, Karimireddy et al.~\cite{karimireddy2021learning} have shown that the CC defender is stable in various $\rho$ settings, from $10^{-1}$ to $10^3$.

\highlight{Frequency Analysis-Based Method (FreqFed)}~\cite{fereidooni2024freqfed}. 
FreqFed is an uninformed defense. 
The defender first applies the discrete cosine transform to the received gradients and gets transformed coefficients. 
The coefficients are then processed with loss-pass filtering to reduce the influence of high-frequency noise. 
Finally, hierarchical density-based spatial clustering (HDBSCAN) is applied to the pairwise cosine distance of the low-frequency coefficients. 
The largest cluster is selected as the benign cluster $\mathcal{S}_{\text{F}}$. 
An aggregated update is obtained by performing FedAvg on the selected cluster, 
\begin{equation}
    f_{\text{FreqFed}}(\update[k]_{\mathcal{C}}) = \frac{1}{|\mathcal{S}_{\text{F}}|}\sum_{b \in \mathcal{S}_{\text{F}}} \update[k,\tau]_{b}. 
\end{equation}

\highlight{SignGuard~\cite{xu2022byzantine}.}
SignGuard is an uninformed defense that utilizes the statistics of gradient signs to filter out malicious updates. 
The defender initially creates a set of indices, $\mathcal{S}_1$, by identifying and excluding outlier gradients. 
Concurrently, another set of indices, $\mathcal{S}_2$, is formed by selecting the largest cluster based on sign agreement. 
The updates are then aggregated using client indices from the intersection $\mathcal{S}_{\text{sg}} = \mathcal{S}_1 \cap \mathcal{S}_2$,
\begin{equation}
    f_{\text{SignGuard}}(\update[k]_{\mathcal{C}}) = \frac{1}{|\mathcal{S}_{\text{sg}} |}\sum_{b \in \mathcal{S}_{\text{sg}}} \update[k,\tau]_{b}. 
\end{equation}

\highlight{Krum and Multi-Krum~\cite{blanchard2017machine}.}
Both Krum and Multi-Krum are informed defenses since they require specifying the parameter $A$, which represents the number of attackers. 
Krum defender chooses one client update from its input that is closest to its neighbors, according to the following operation:
\begin{equation}\label{eq:krum}
    f_{\text{Krum}}(\update[k]_{\mathcal{C}}; A) = \underset{\update[k,\tau]_{i}}{\operatorname{argmin}} \sum_{i \rightarrow j}\left\|\update[k,\tau]_i-\update[k,\tau]_j\right\|^2,
\end{equation}
where $i \rightarrow j$ is the indices of the $M-A-2$ nearest neighbors of $\update[k,\tau]_{i}$ based on the Euclidean distance.
Multi-Krum extends Krum by selecting $c$ model updates and averages selected updates. 
Specifically, Multi-Krum performs Krum in~\eqref{eq:krum} $c$ times, each time selecting an update and moving it from the received update set $\update[k]_{\mathcal{C}}$ to the Multi-Krum candidate set.

\highlight{Divide and Conquer (DnC)~\cite{shejwalkar2021manipulating}.}
DnC is an informed defense that assumes knowledge of the number of attackers, $A$.
DnC defender first randomly selects a set of indices of coordinates to sparsify/subsample the gradients, keeping $s$ valid entries out of $d$.
The defender then constructs an $M \times s$ matrix $\mG$ by concatenating subsampled gradients and normalizing them to $\tilde{\mG}$ by subtracting the mean gradients. 
DnC detects the attackers by projecting these centered gradients in $\tilde{\mG}$ along their principal right singular eigenvector $\bm{v}$ and determining outlier scores. 
Given a filtering factor $c \in (0,1)$, gradients with the lowest outlier scores $M - c A$ are selected as benign updates.
Thus, a benign index set $\mathcal{S}_{\text{D}}$ can be constructed.  
The final update is computed by averaging the selected updates,
\begin{equation}
    f_{\text{DnC}}(\update[k]_{\mathcal{C}}; A) = \frac{1}{|\mathcal{S}_{\text{D}}|}\sum_{b \in \mathcal{S}_{\text{D}}} \update[k,\tau]_{b}. 
\end{equation}

\highlight{Trimmed Mean (TM)~\cite{yin2018byzantine}.}
TM is considered an informed defense. 
It calculates the mean after excluding a certain percentage of the highest and lowest values for each dimension of the gradient vectors.
This defense requires specifying the parameter $A$, which indicates the number of attackers.

\subsection{Surveys on Byzantine Attacks and Defenses}
Multiple surveys~\cite{lyu2022privacy, shi2022challenges} have summarized the challenges and solutions in Byzantine robust FL.   
However, a clear evaluation of the efficacy and limitations of various defenses remains elusive. 
This gap has resulted in ambiguity when practitioners seek a general-purpose aggregation rule during deployment.
From an experimental standpoint, studies by Shejwalkar et al.~\cite{shejwalkar2022back}, Han et al.~\cite{han2023fedmlsecurity}, and Li et al.~\cite{li2023experimental, li2024blades} have investigated the effectiveness of defenses and attacks in FL across various settings, including the cross-silo setting with both IID and non-IID data distributions~\cite{li2023experimental, li2024blades}, cross-device production-level FL~\cite{shejwalkar2022back}, and for large language models~\cite{han2023fedmlsecurity}. 

Generally speaking, the arms race between attackers and defenders is influenced by many factors, including learning tasks and hyperparameters~\cite{li2024blades}.
Defenders aim to perform well on average while minimizing exposure to worst-case vulnerabilities. 
In parallel, attackers may alter their tactics upon recognizing the ineffectiveness of their initial approach. 
The fraction of Byzantine attacks may also vary in different communication rounds. 
To the best of our knowledge, the effect of attackers altering tactics is not well explored in the literature.For example, most defenses have been evaluated under fixed attack strategies that remain unchanged throughout the FL training.   
Furthermore, there is no universal aggregation algorithm suggested in previous studies. 
In this work, we examine the attack-agnostic setting from both the attacker's side and the defender's side. 
Specifically, from the attacker’s side, we examine the potential increase in attack impact when multiple tactics are allowed during FL training. 
From the defender's side, we propose a general-purpose aggregation rule that performs the best under such attack-agnostic settings.

\section{Hybrid Defense For Attack-Agnostic FL}\label{sec:hybrid_defense}
In this section, we present the feasibility and necessity of designing hybrid defenses in attack-agnostic settings. 
We first stand on the defender's side and aim to identify a general-purpose aggregation strategy.  
As we have reviewed in Section~\ref{section:preliminaries}, attackers can switch tactics mid-way through training.
Defenders need to be as flexible as attackers.
Some defenses work well against specific attacks, but might not be effective across the board. 
If the defender is empowered with various defenses, robustness can be potentially improved.  
We propose a hybrid defense framework that accommodates two practical scenarios.

\highlight{Reference Dataset Available.} 
When an evaluation dataset $\mathcal{D}_0$ is available on the server, it can be used as a reference to dynamically select the most appropriate defense strategy at each communication round.
The use of evaluation datasets has been extensively investigated in prior works, and the effectiveness has been verified even when the size is in the range of hundreds~\cite{xie2020zeno++, cao2021fltrust, fang2024byzantine, yan2025skymask}. 
Such datasets may be publicly available~\cite{zhao2018federated}, contributed by volunteer employees of the service provider~\cite{cao2021fltrust}, or intrinsically held by a node in a decentralized manner~\cite{fang2024byzantine}. 
Consider a set of defense mechanisms $\mathcal{M} = \{f_1, f_2, \ldots, f_{D}\}$, where each $f_j$ represents a robust aggregation function.
The server selects the aggregation function that minimizes the empirical risk.
Formally, the aggregation function $f^{(k)}$ at round $k$ is determined by:
\begin{equation} \label{eq:hybrid1}
    f^{(k)}_{\text{Hybrid-R}} = \argmin_{f_j \in \mathcal{M}}\, R\left[ \w[k] - \eta_{g} f_j(\update[k,\tau]_{\mathcal{C}}; \mathcal{I}_{j}); \mathcal{D}_{0}\right], 
\end{equation}
where the subscript ``Hybrid-R'' denotes the defense based on a reference dataset available.

\highlight{Reference Dataset Unavailable.} 
When an evaluation dataset is unavailable, there are multiple ways to design a hybrid aggregation scheme based on a defense set $\mathcal{M}$. 
The first option is to apply cascaded robust aggregation rules.
In this framework, updates are progressively filtered through multiple defenses.
This is analogous to cascaded filtering in signal processing, where noise is reduced incrementally through successive stages. 
Formally, the hybrid defense may be designed as follows:
\begin{subequations}
\begin{align}
    \update[k]_{\mathcal{H}} & := \{ f_{i}(\update[k]_{\mathcal{C}}; \mathcal{I}_{i}) \,\mid\, f_i \in \mathcal{M} \}, \\
    \hupdate[k]_{\text{Hybrid-NR}} & := f_{\text{final}}(\update[k]_{\mathcal{H}} ; \mathcal{I}_{\text{final}}),
\end{align}
\end{subequations}
where the subscript ``Hybrid-NR'' denotes the defense that does not require a reference dataset and $f_{\text{final}}$ denotes the aggregation function at the final stage. 
This formulation introduces flexibility, as the choice of function $f_{\text{final}}$ may affect the final outcome.
For example, a strategy that begins with Krum defense in $\mathcal{M}$ to remove large-norm updates, followed by SignGuard to enforce sign consistency, might yield different results than the reverse order. 
While traditional cascaded filtering in signal processing can often be described by multiplication of frequency responses, making the filter order theoretically inconsequential, the Hybrid-NR defense we employ here is different. 
Specifically, robust aggregation functions in FL are inherently nonlinear. 
Thus, the sequence in which these aggregation methods are applied significantly influences the final outcome. 
Meanwhile, in cascaded filtering, the input signal sequentially passes through multiple filters, each designed to capture or eliminate different types of noise or disturbance. Each subsequent filter receives a cleaner version of the signal, progressively refining its quality. Similarly, in the Hybrid-NR defense, each robust aggregation rule acts as a filter, removing suspicious updates or features from the aggregated set. 
This layered approach helps ensure that updates that pass the final aggregation stage have already undergone rigorous scrutiny by preceding methods, thus increasing robustness and resilience against diverse adversarial tactics.
For Hybrid-NR defense in this work, we consider FreqFed defense as $f_{\text{final}}$ unless otherwise specified.

There are many other options when it comes to designing a hybrid defense. 
For example, the server may rely on the intersection of indices suggested by multiple robust aggregation schemes or criteria, such as $\set_{1} \cap \set_{2}$ designed in SignGuard~\cite{xu2022byzantine}. 
This idea has been explored in some related works~\cite{krauss2023avoid, krauss2023mesas}. 
By aggregating only updates that appear at the intersection of these subsets, the server incorporates opinions from multiple ``expert" defenses, which can improve robustness against diverse attack strategies. 
Although this approach benefits from leveraging agreement among different defenses, it also comes with inherent limitations. 
For example, the intersection may be empty in the case of a high disagreement between the strategies. 
Due to the high heterogeneity in FL and empirical observations of frequent disagreement among robust aggregation schemes, this type of intersection-based defense is not included in this study.

\begin{figure*}[!tb]
    \begin{overpic}[width=\linewidth, height=0.5\linewidth]{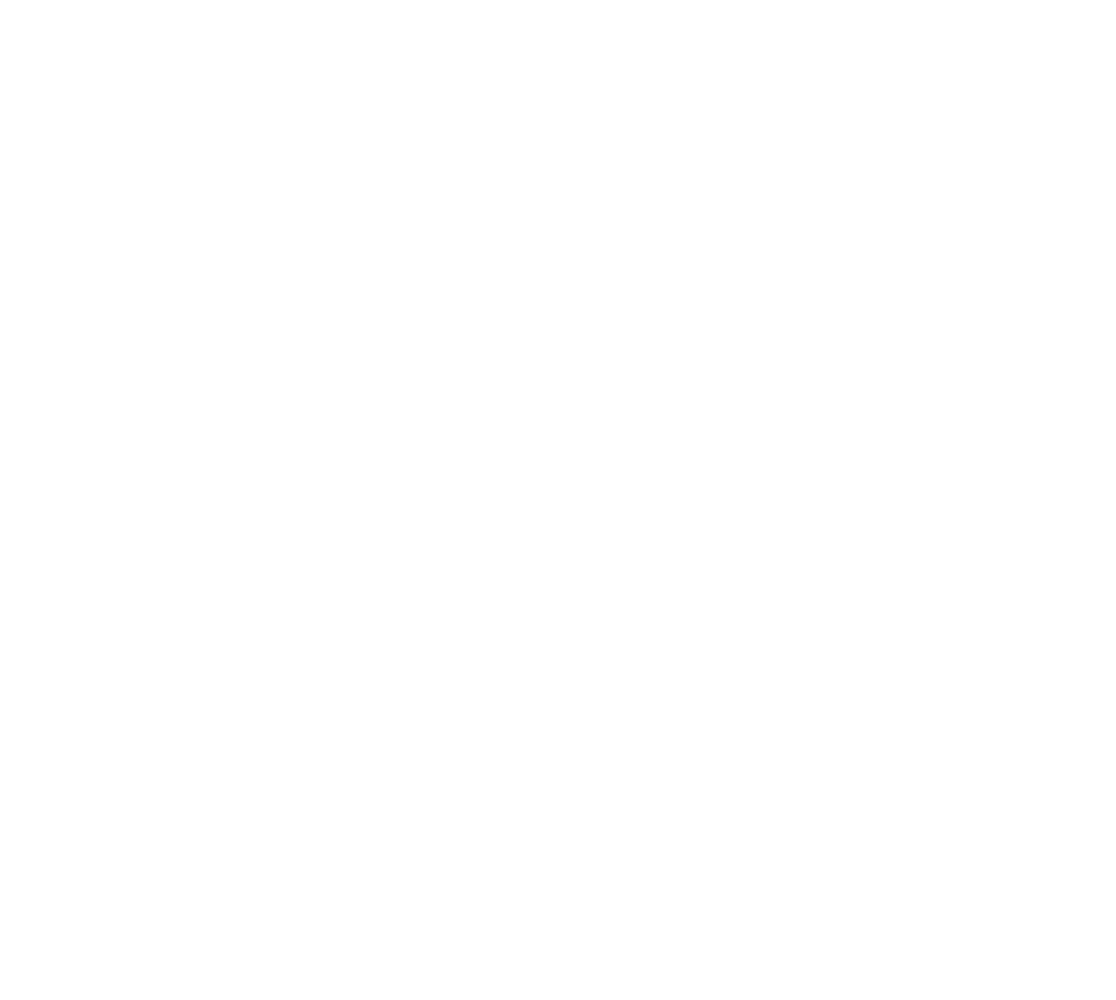}
    \put(12, -1){\includegraphics[width=0.8\linewidth]{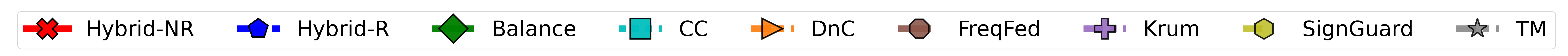}}
    \put(0, 36){\rotatebox{90}{\intab{\bf UCI-HAR}}}
    \put(0, 21){\rotatebox{90}{\intab{\bf F-MNIST}}}
    \put(0, 6){\rotatebox{90}{\intab{\bf CIFAR-10}}}
    
    \put(8, 48){\intab{\bf 3DFed }}
    \put(24, 48){\intab{\bf IPM ($\epsilon = 1$) }}
    \put(40, 48){\intab{\bf Min-Max}}
    \put(58, 48){\intab{\bf NT }}
    \put(74, 48){\intab{\bf ROP }}
    \put(91, 48){\intab{\bf SF }}
    
    \put(82, 33){\includegraphics[width=0.18\linewidth]{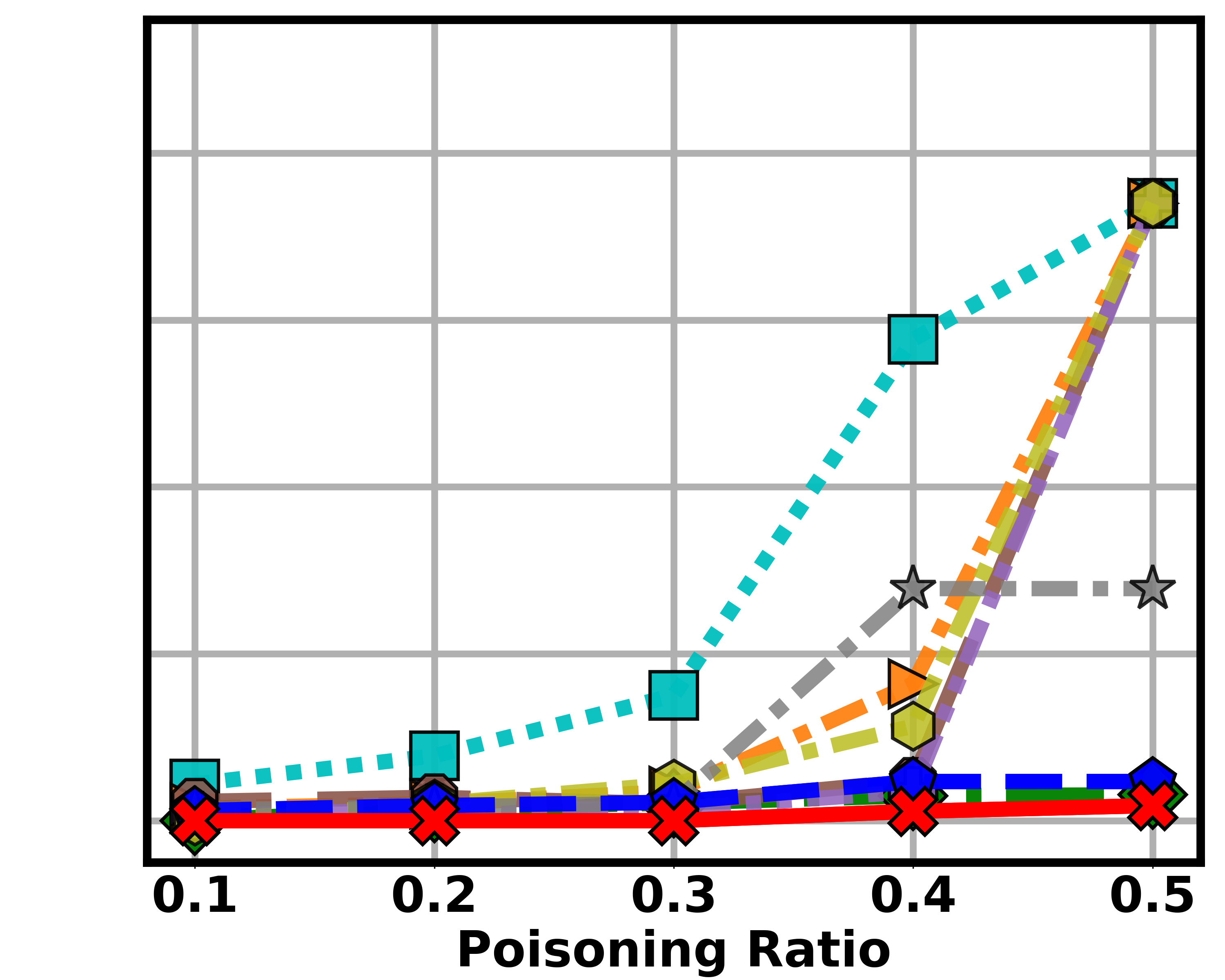}}
    \put(65.9, 33){\includegraphics[width=0.18\linewidth]{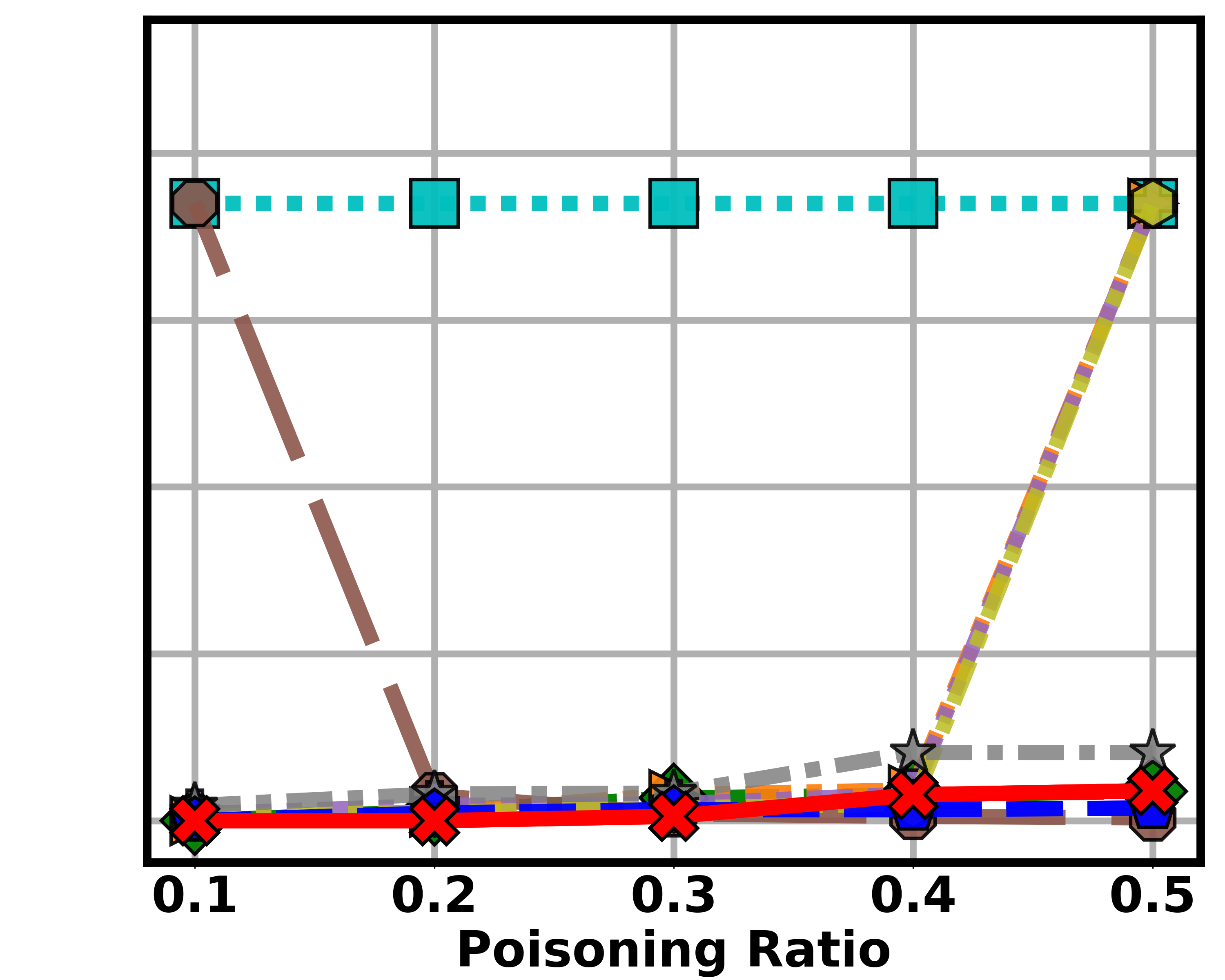}}
    \put(49.8, 33){\includegraphics[width=0.18\linewidth]{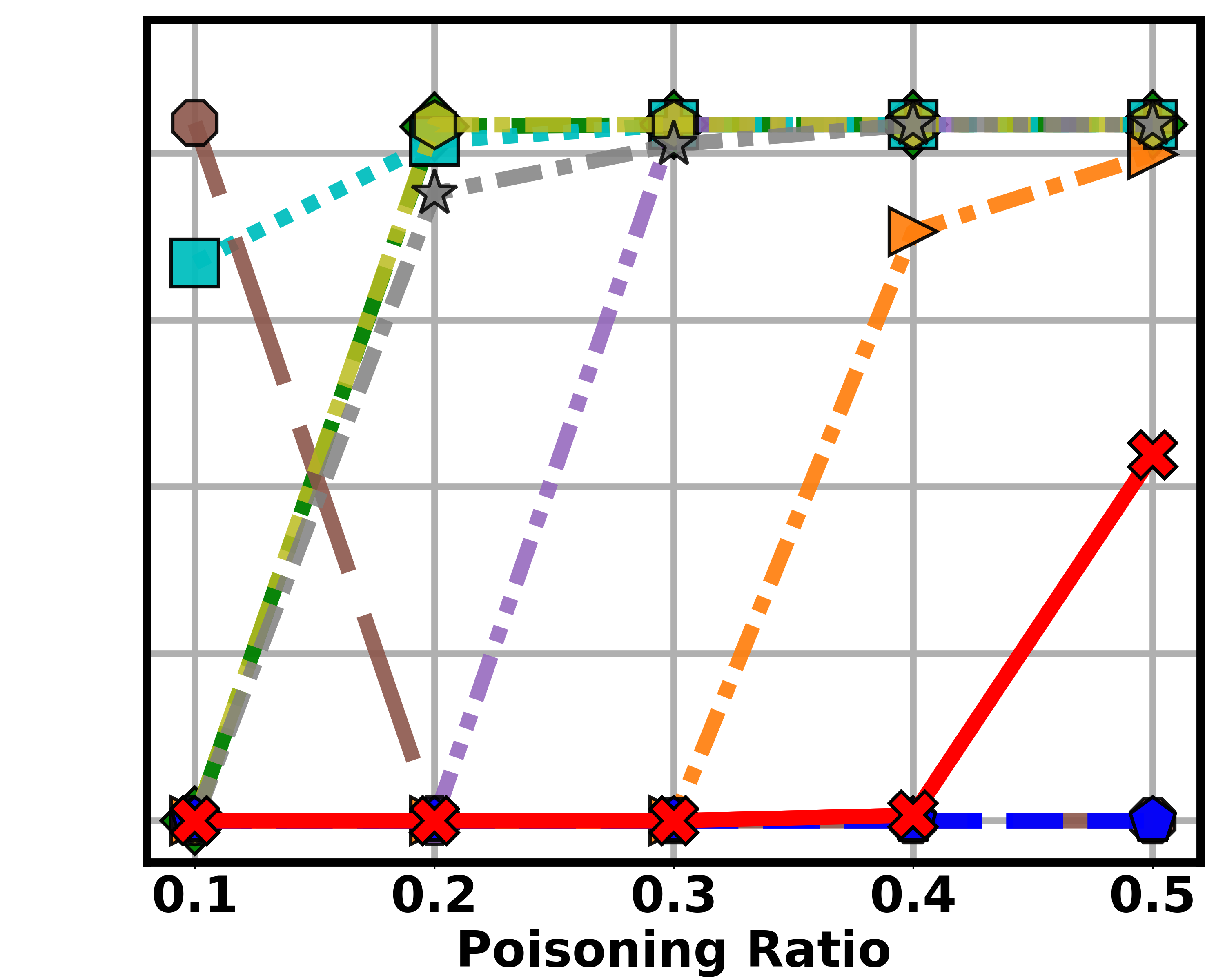}}
    \put(33.7, 33){\includegraphics[width=0.18\linewidth]{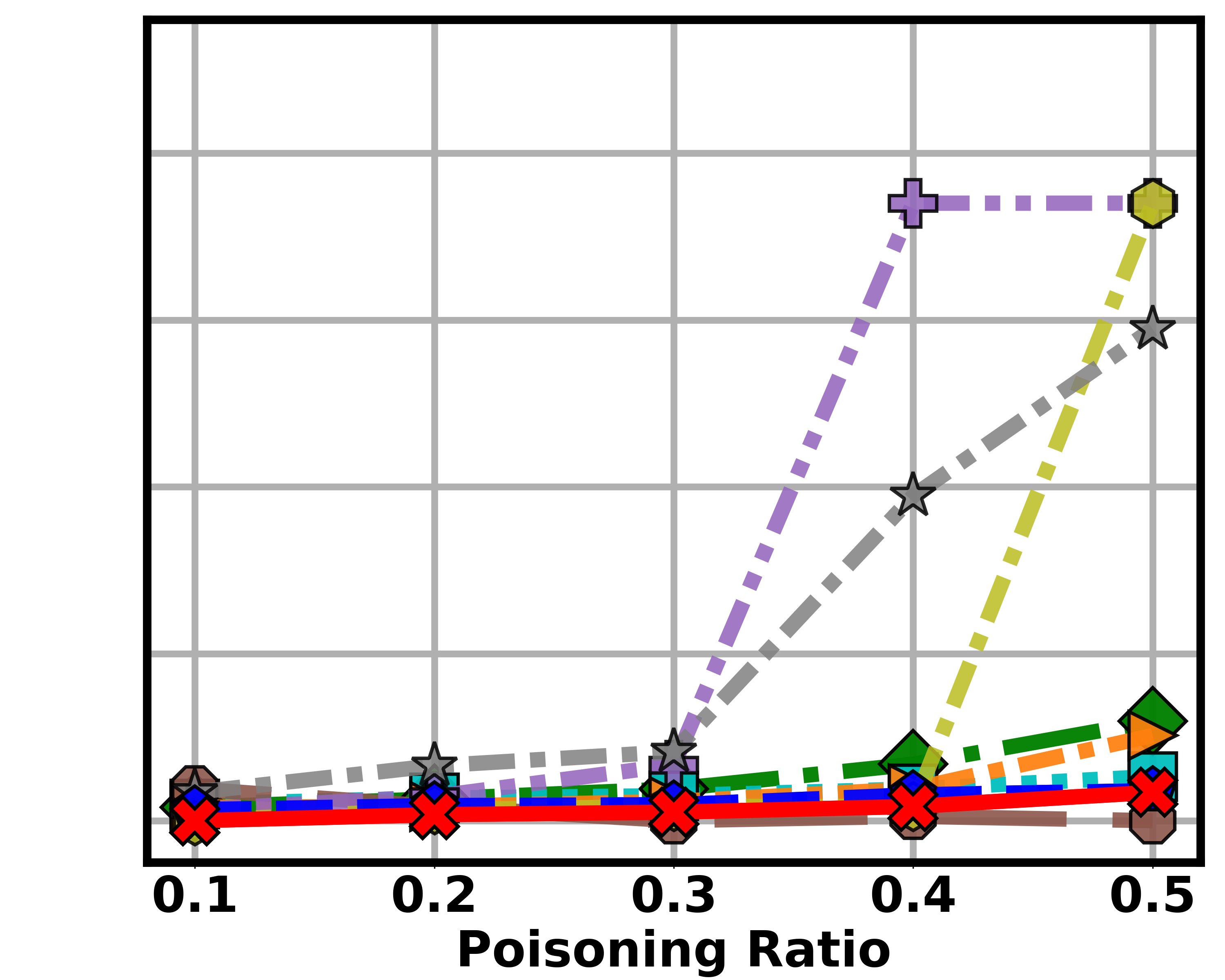}}
    \put(17.6, 33){\includegraphics[width=0.18\linewidth]{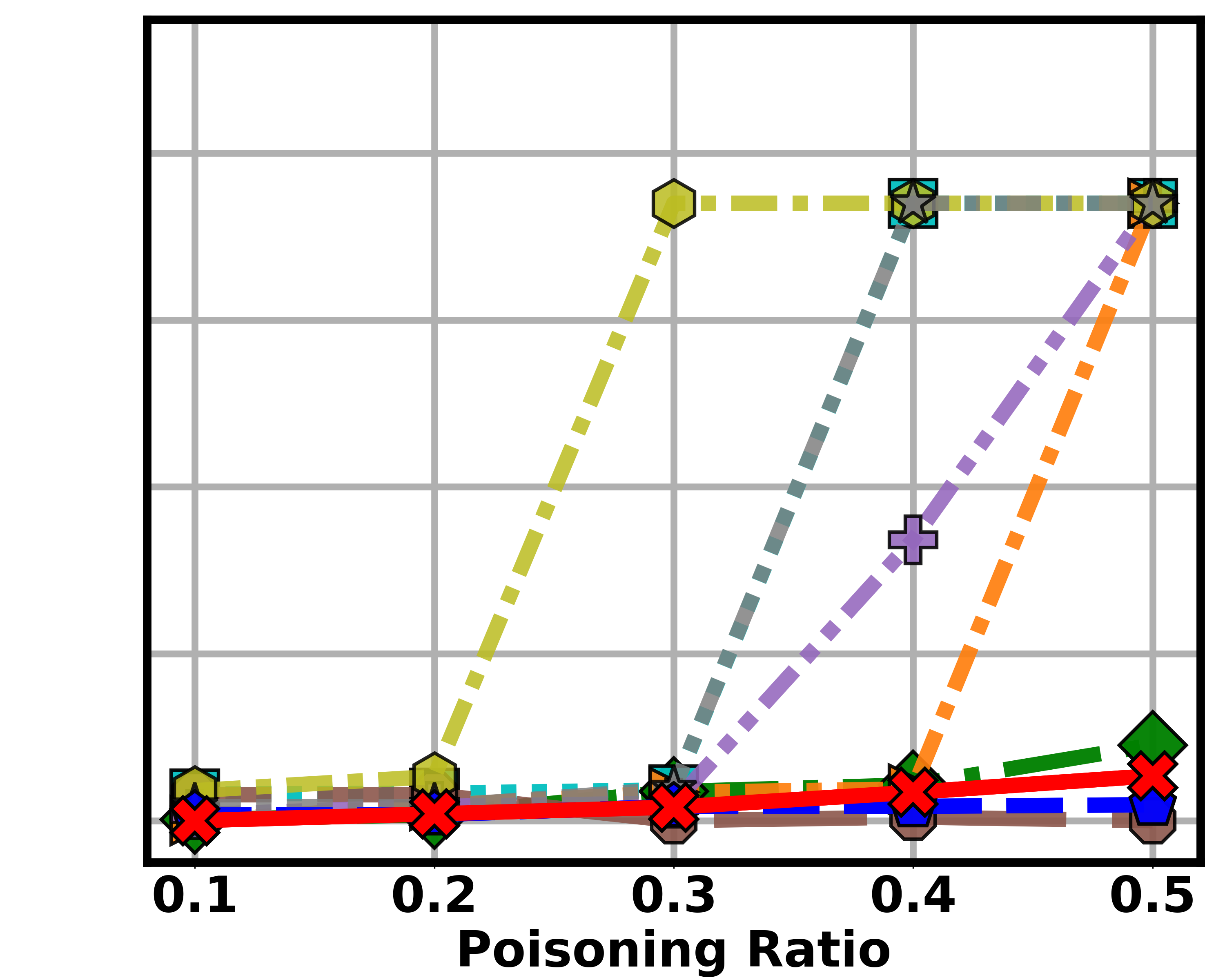}}
    \put(1.5, 33){\includegraphics[width=0.18\linewidth]{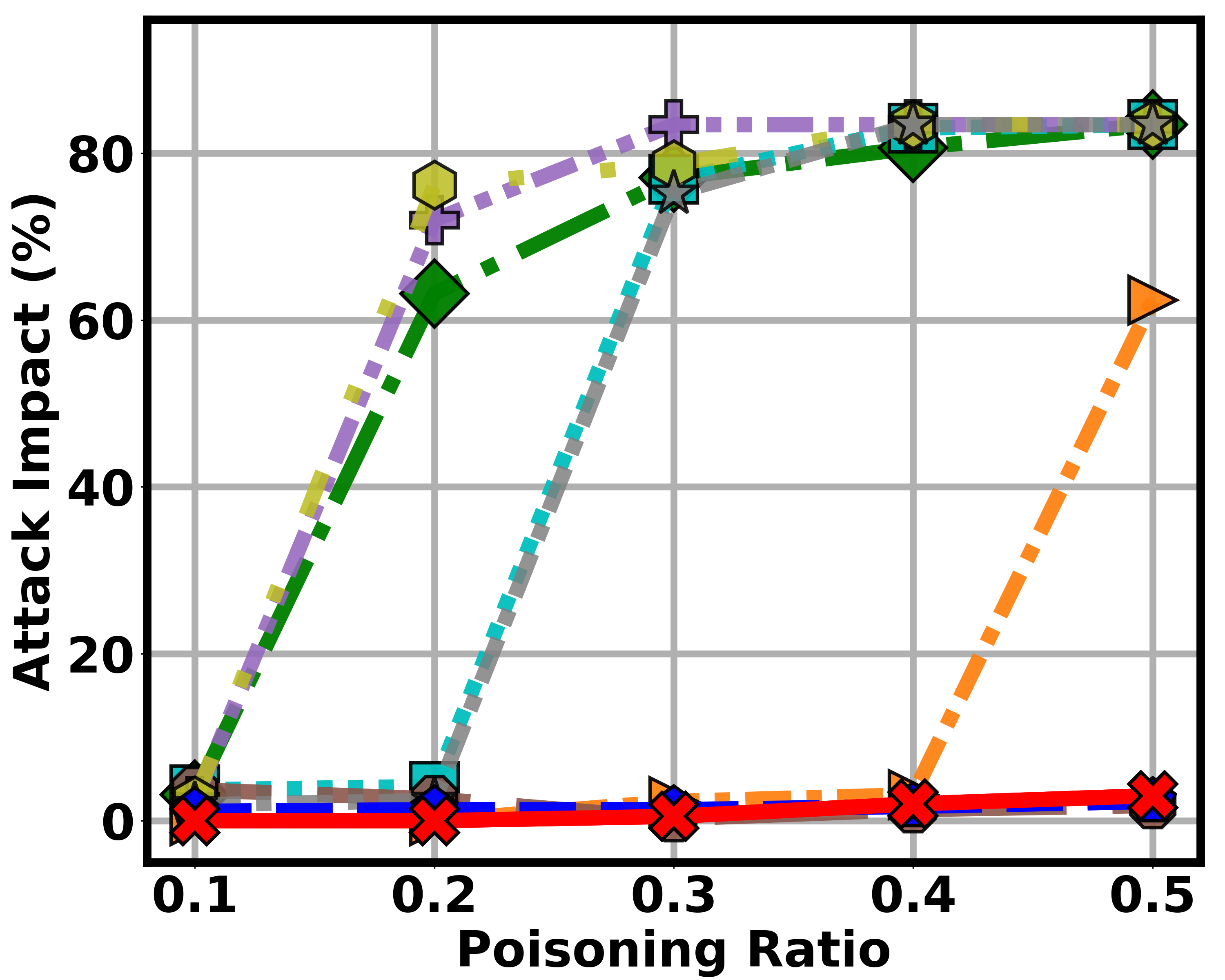}}
    
    \put(82, 18){\includegraphics[width=0.18\linewidth]{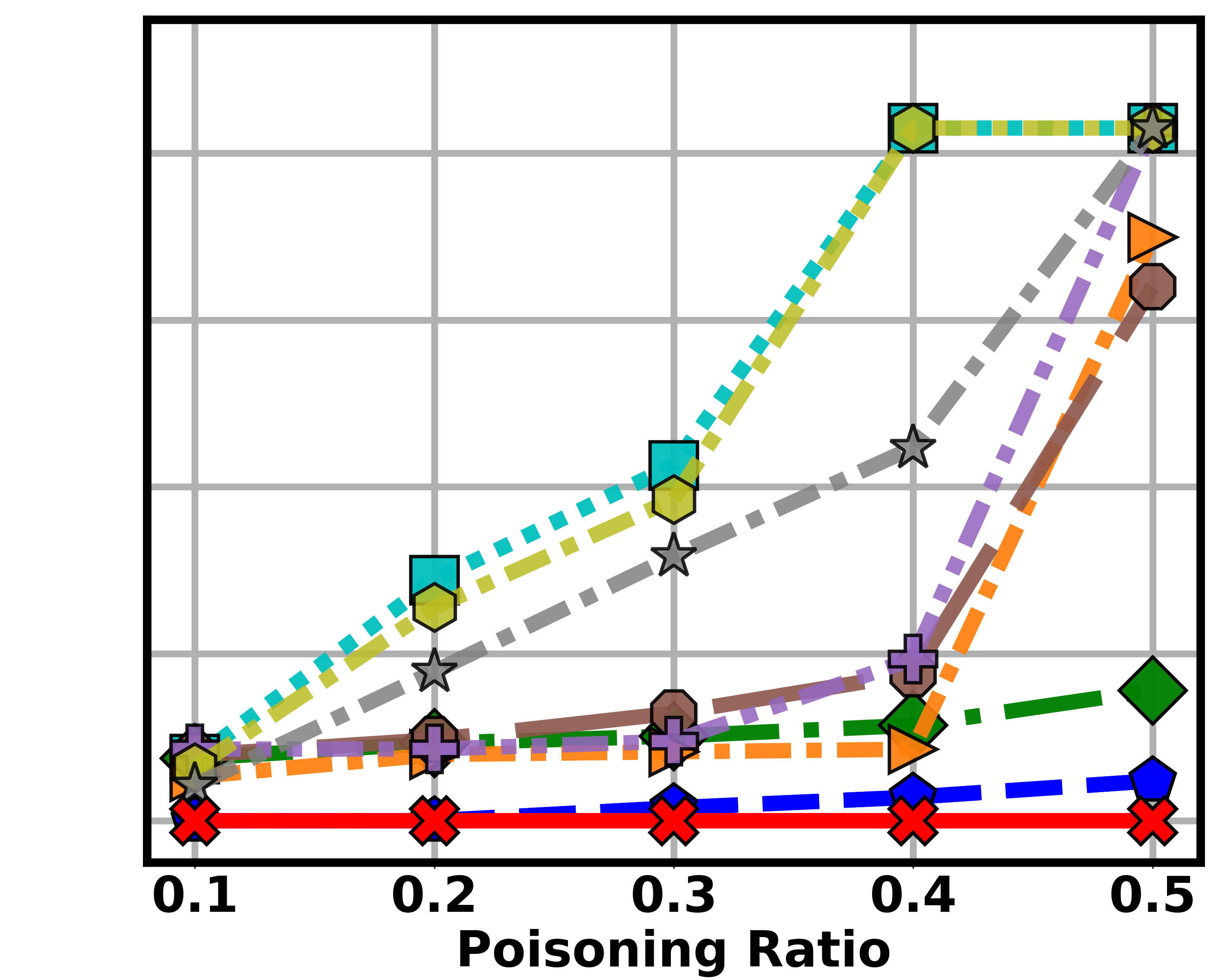}}
    \put(65.9, 18){\includegraphics[width=0.18\linewidth]{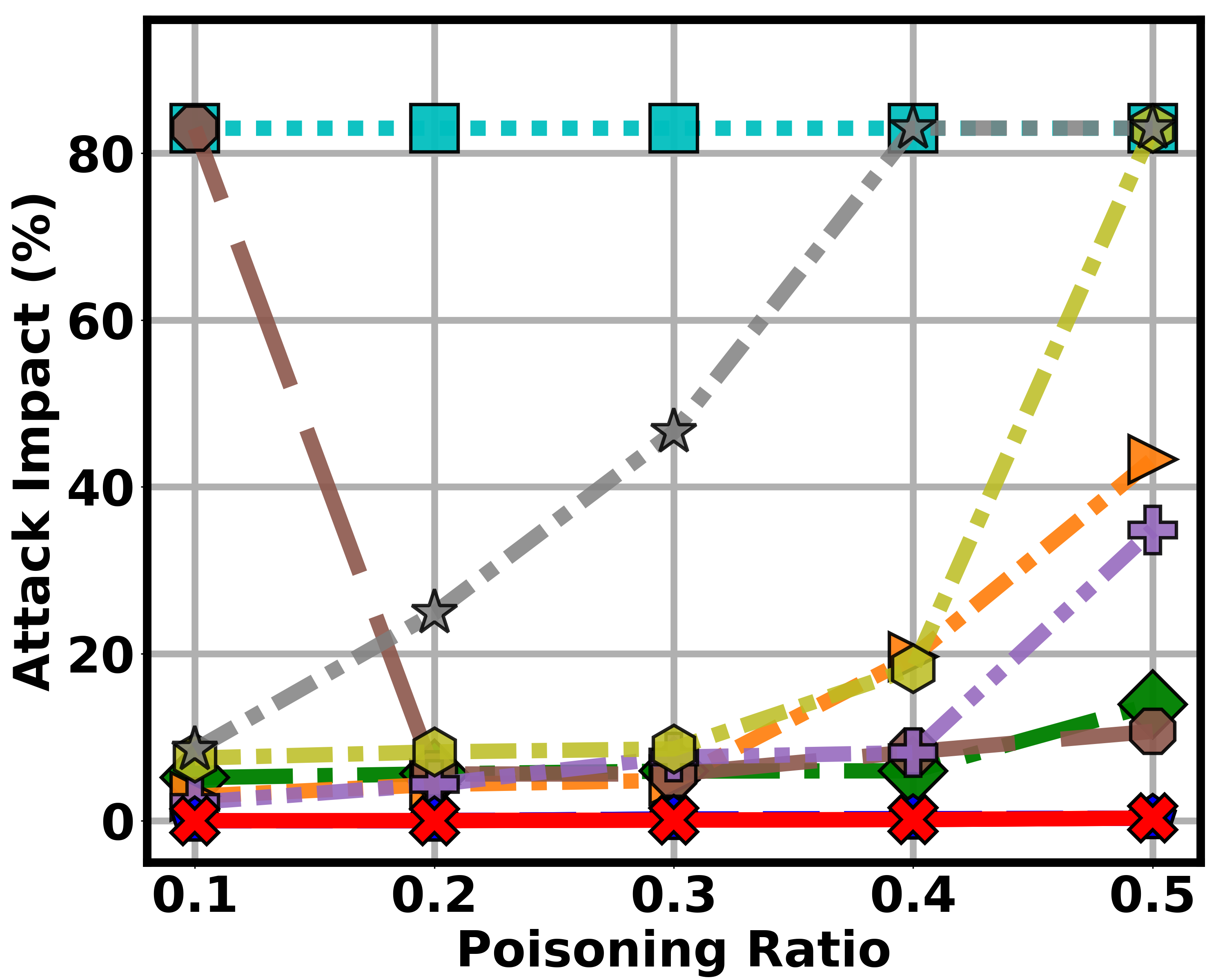}}
    \put(49.8, 18){\includegraphics[width=0.18\linewidth]{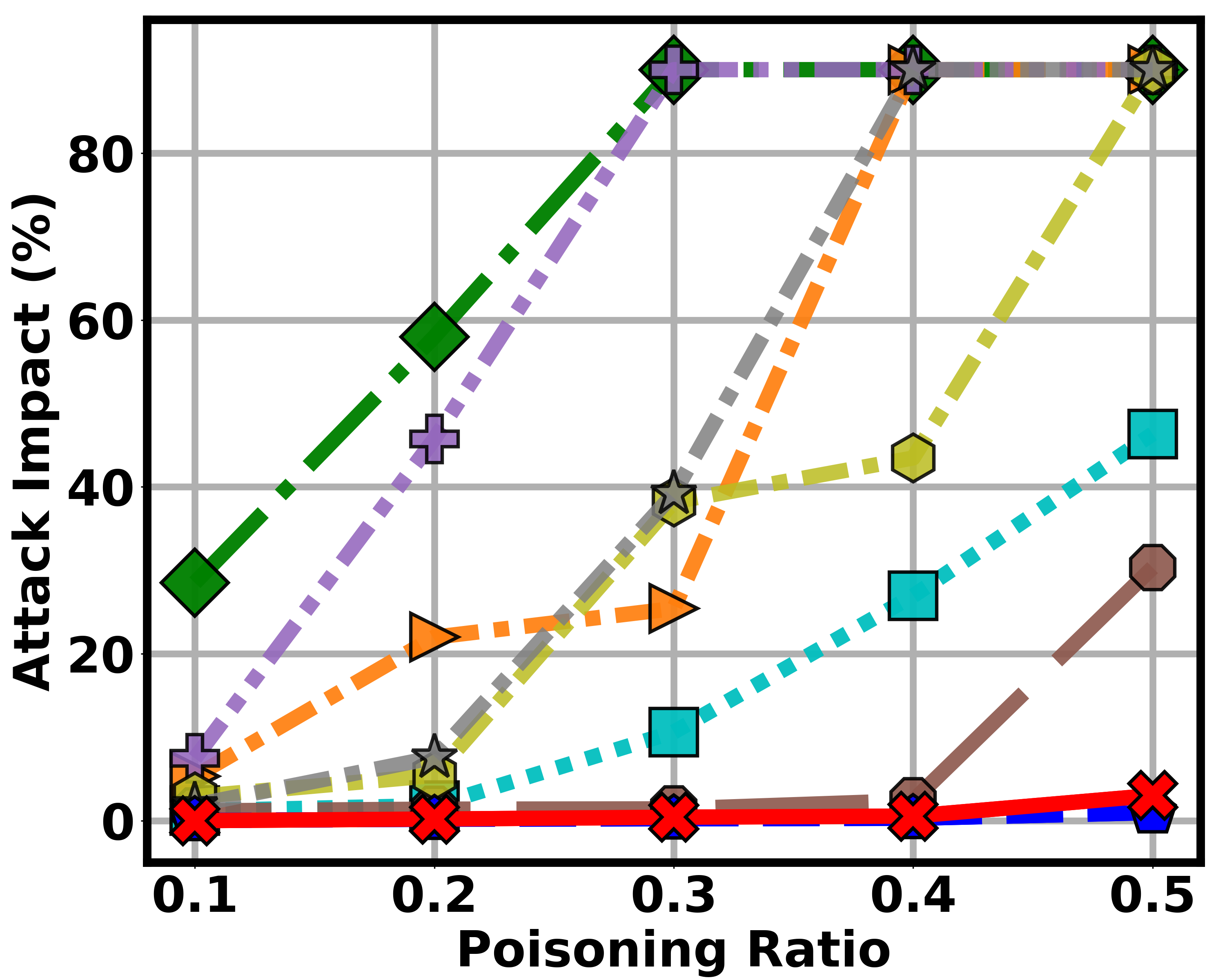}}
    \put(33.7, 18){\includegraphics[width=0.18\linewidth]{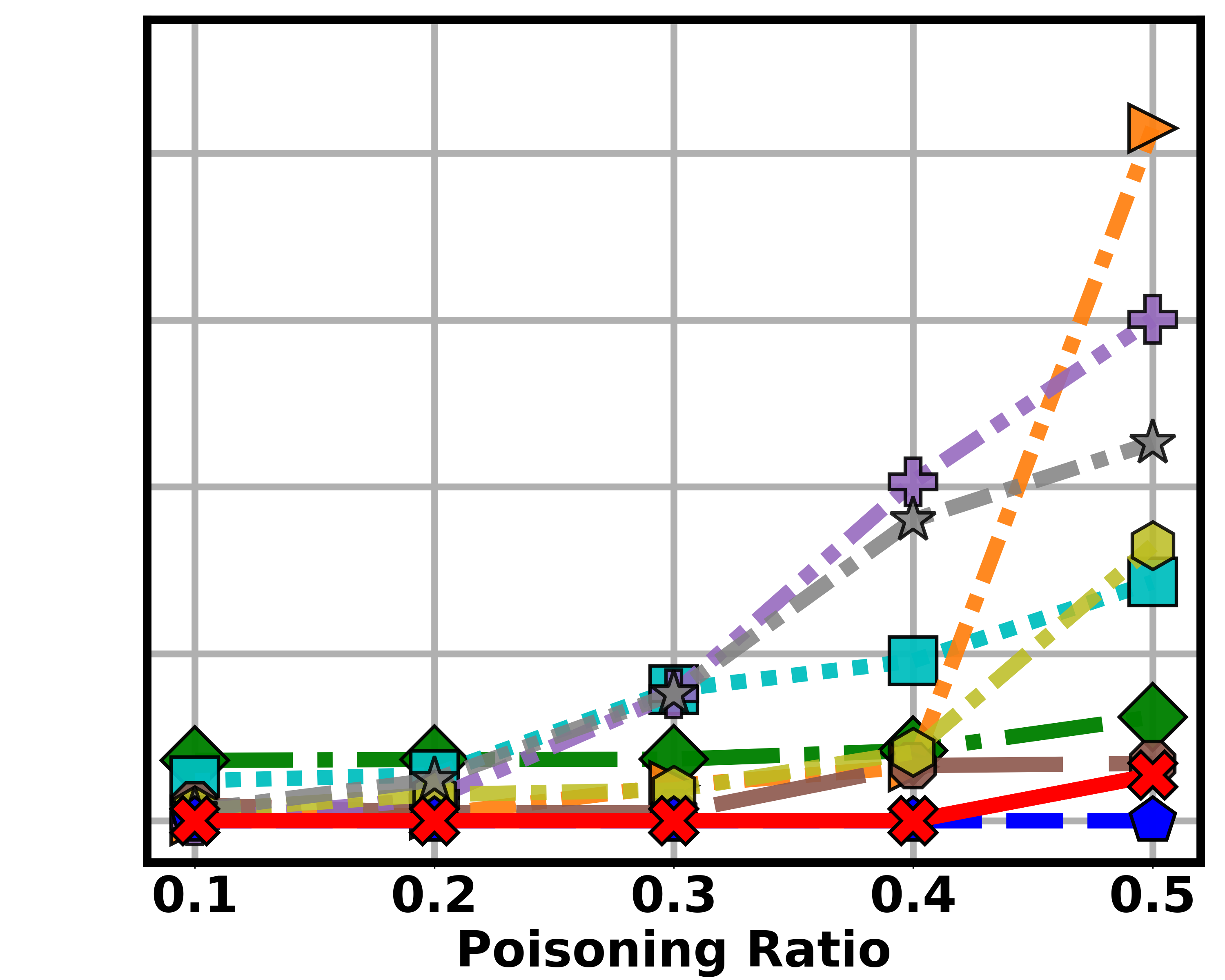}}
    \put(17.6, 18){\includegraphics[width=0.18\linewidth]{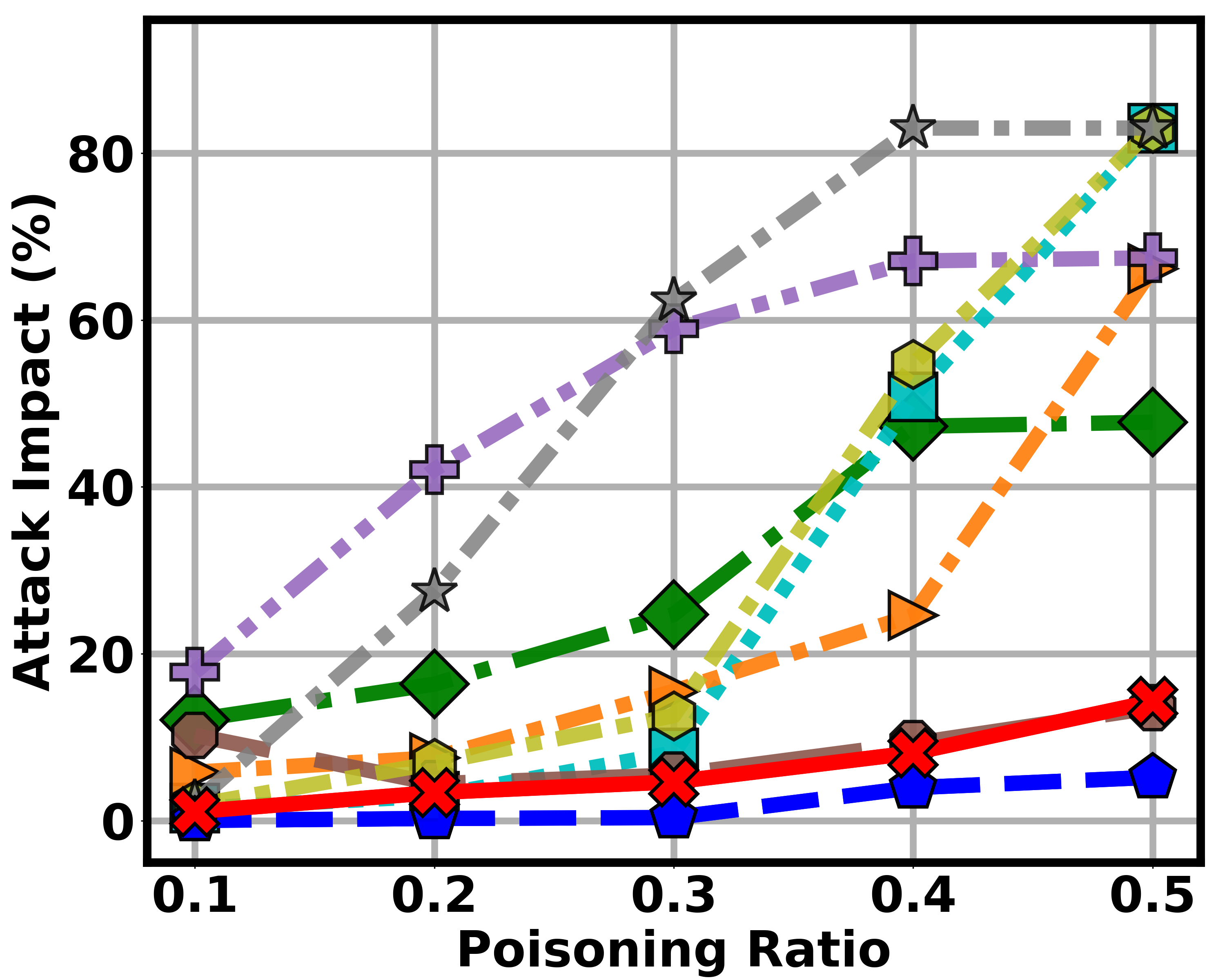}}
    \put(1.5, 18){\includegraphics[width=0.18\linewidth]{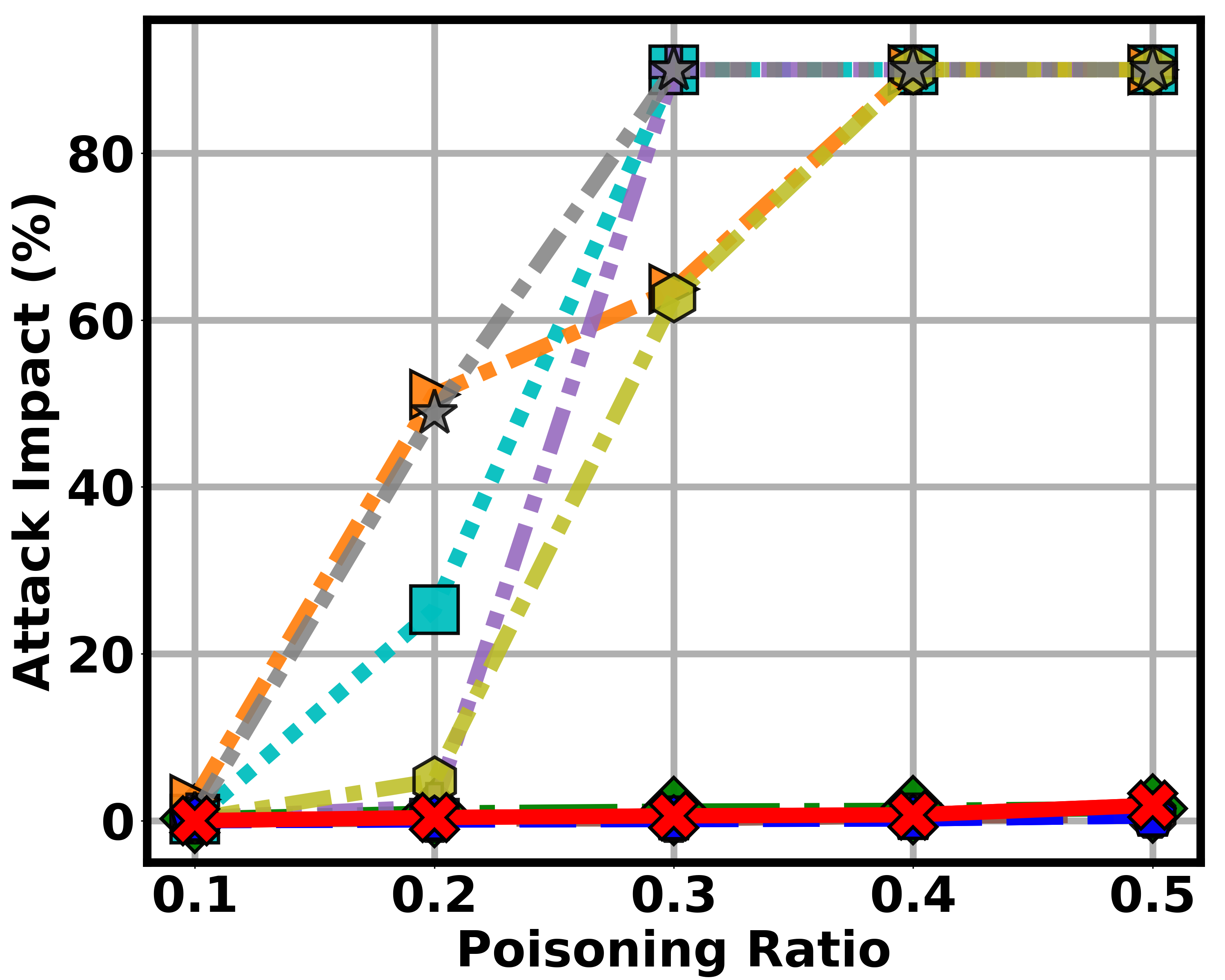}}

    \put(82, 3){\includegraphics[width=0.18\linewidth]{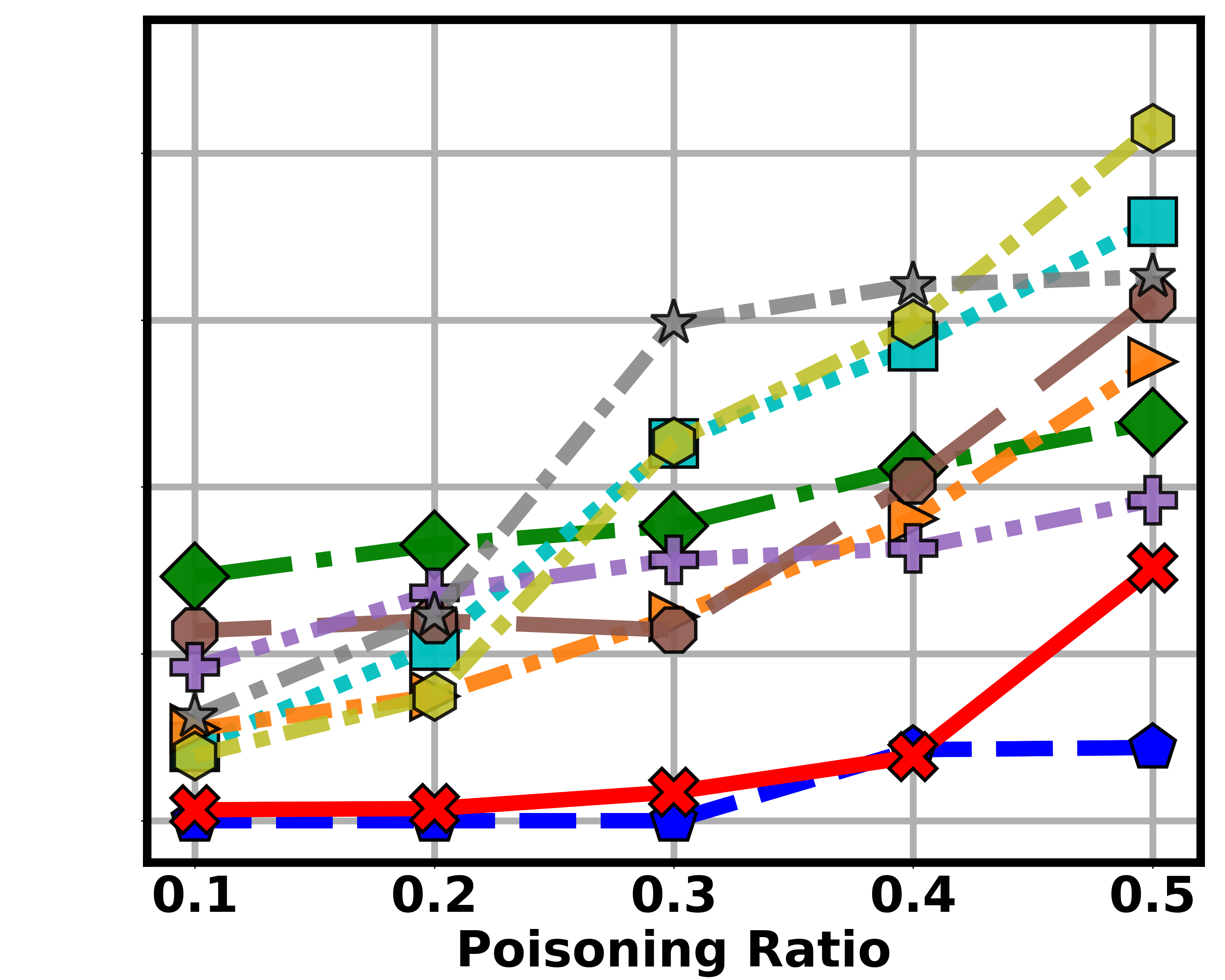}}
    \put(65.9, 3){\includegraphics[width=0.18\linewidth]{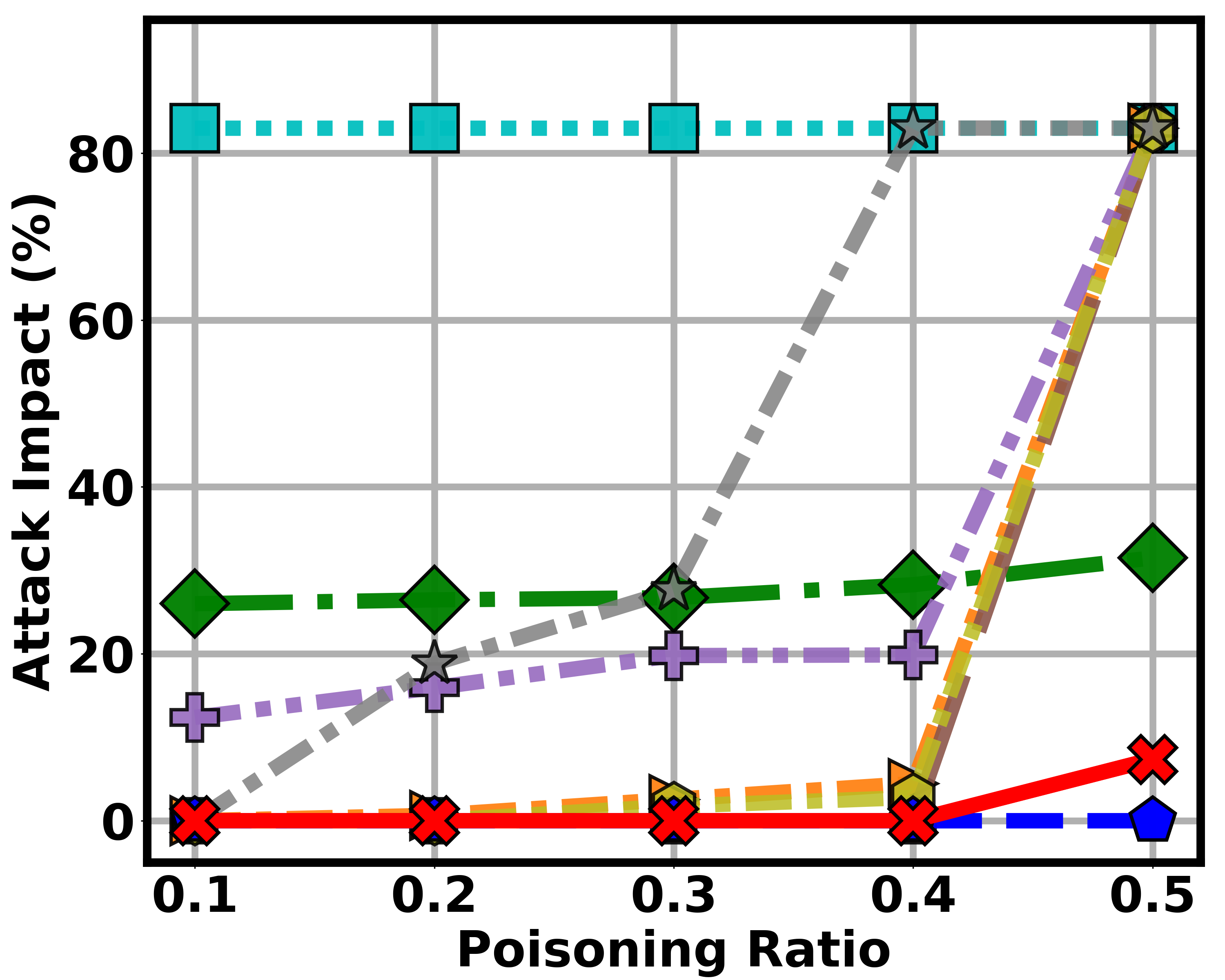}}
    \put(49.8, 3){\includegraphics[width=0.18\linewidth]{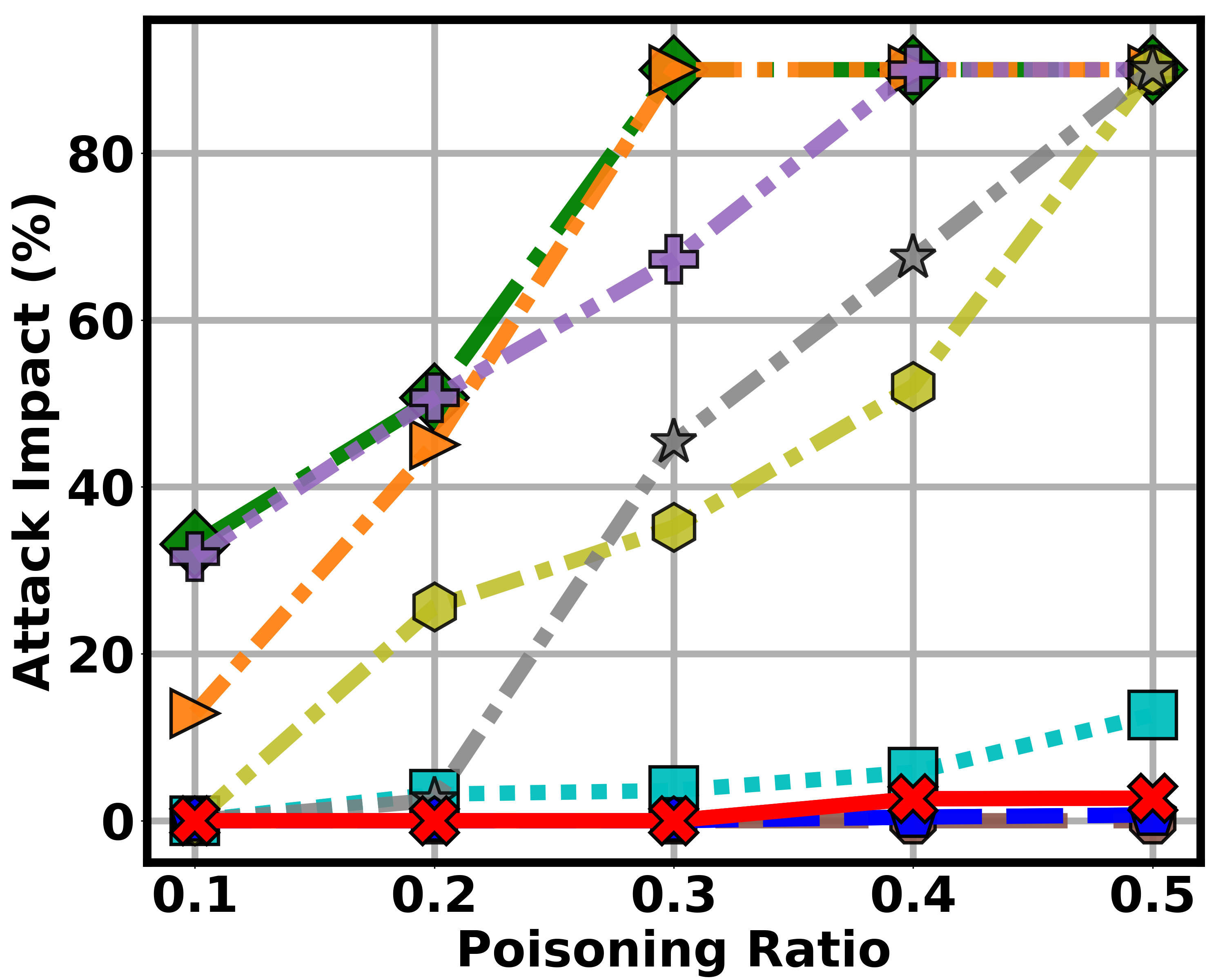}}
    \put(33.7, 3){\includegraphics[width=0.18\linewidth]{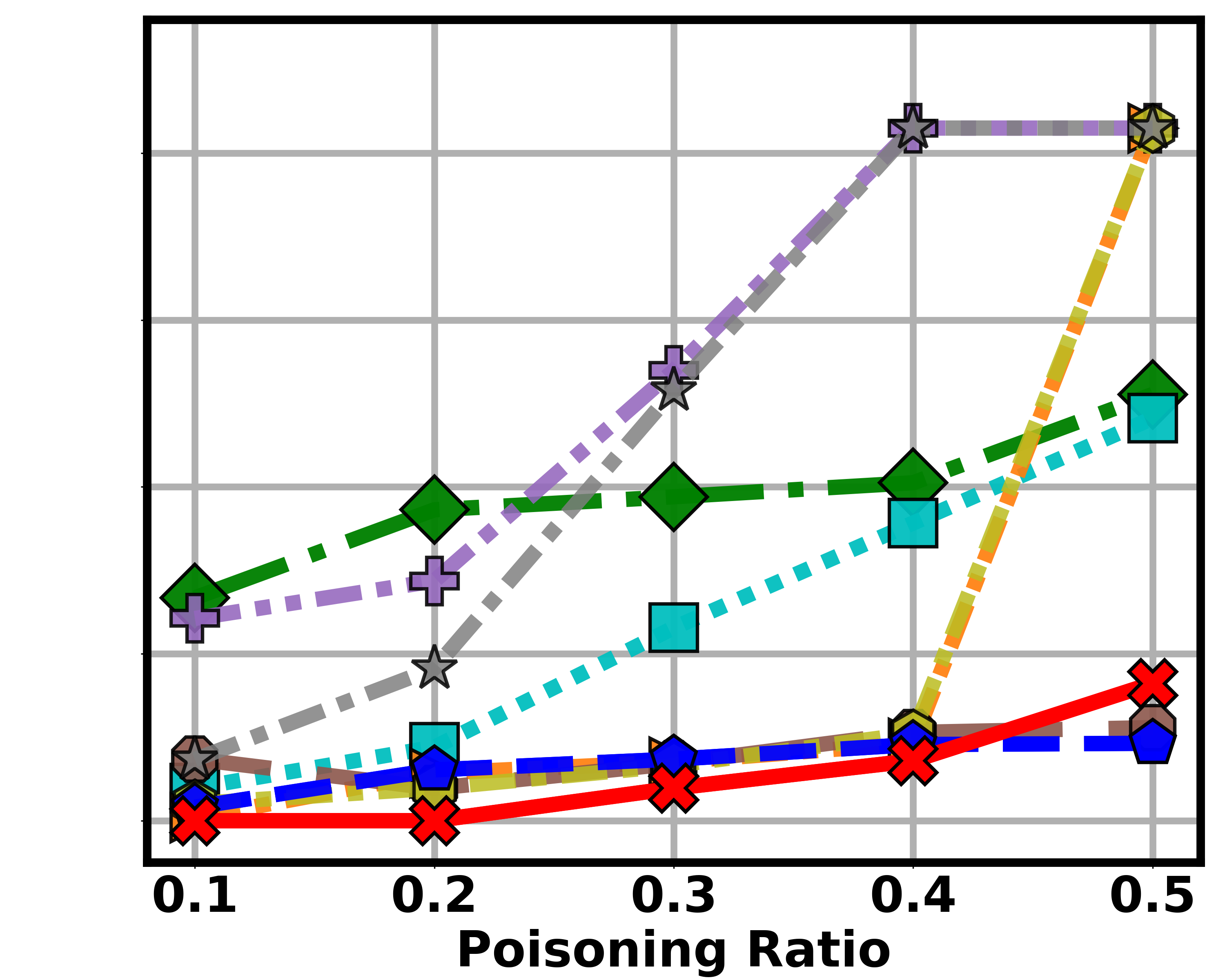}}
    \put(17.6, 3){\includegraphics[width=0.18\linewidth]{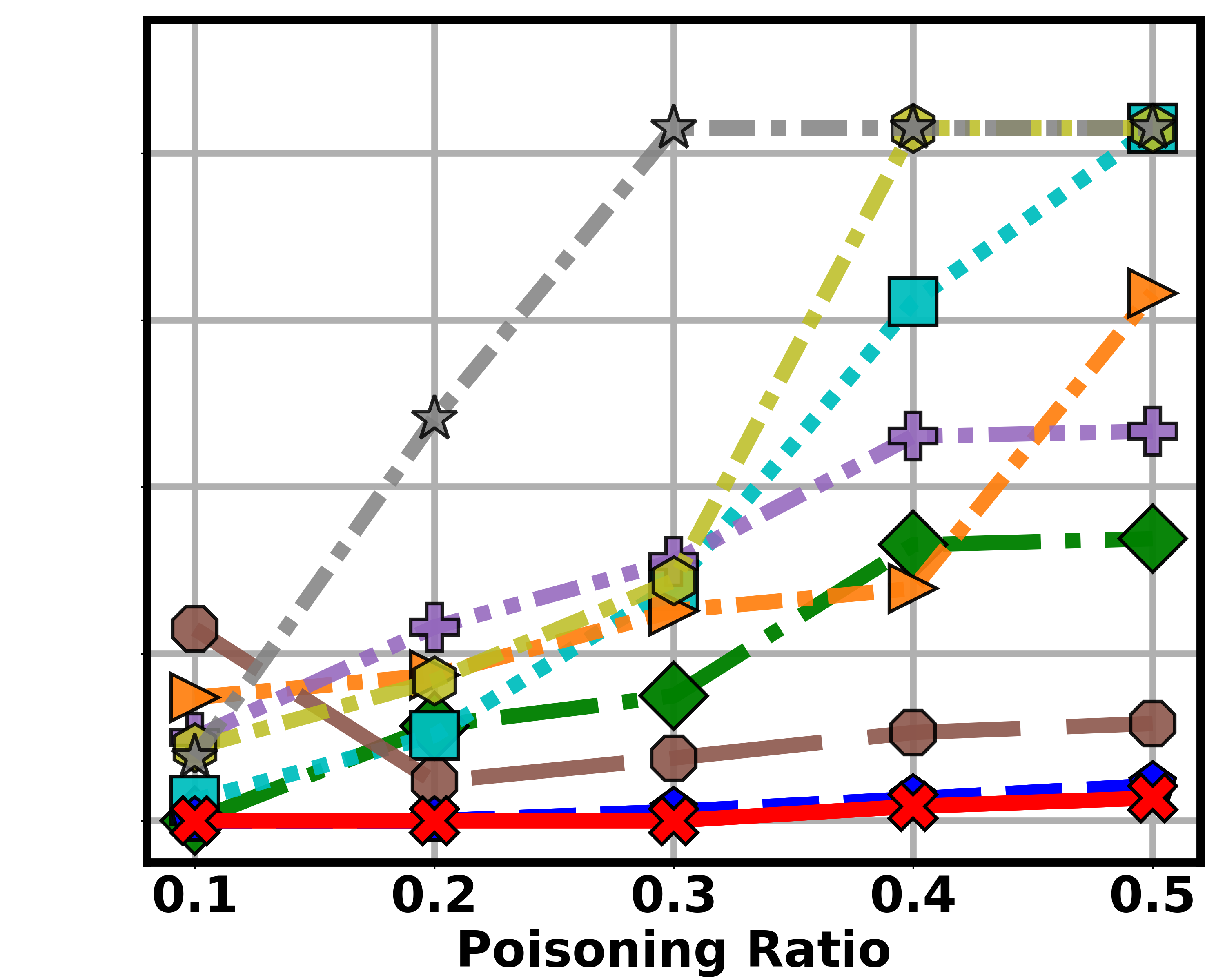}}
    \put(1.5, 3){\includegraphics[width=0.18\linewidth]{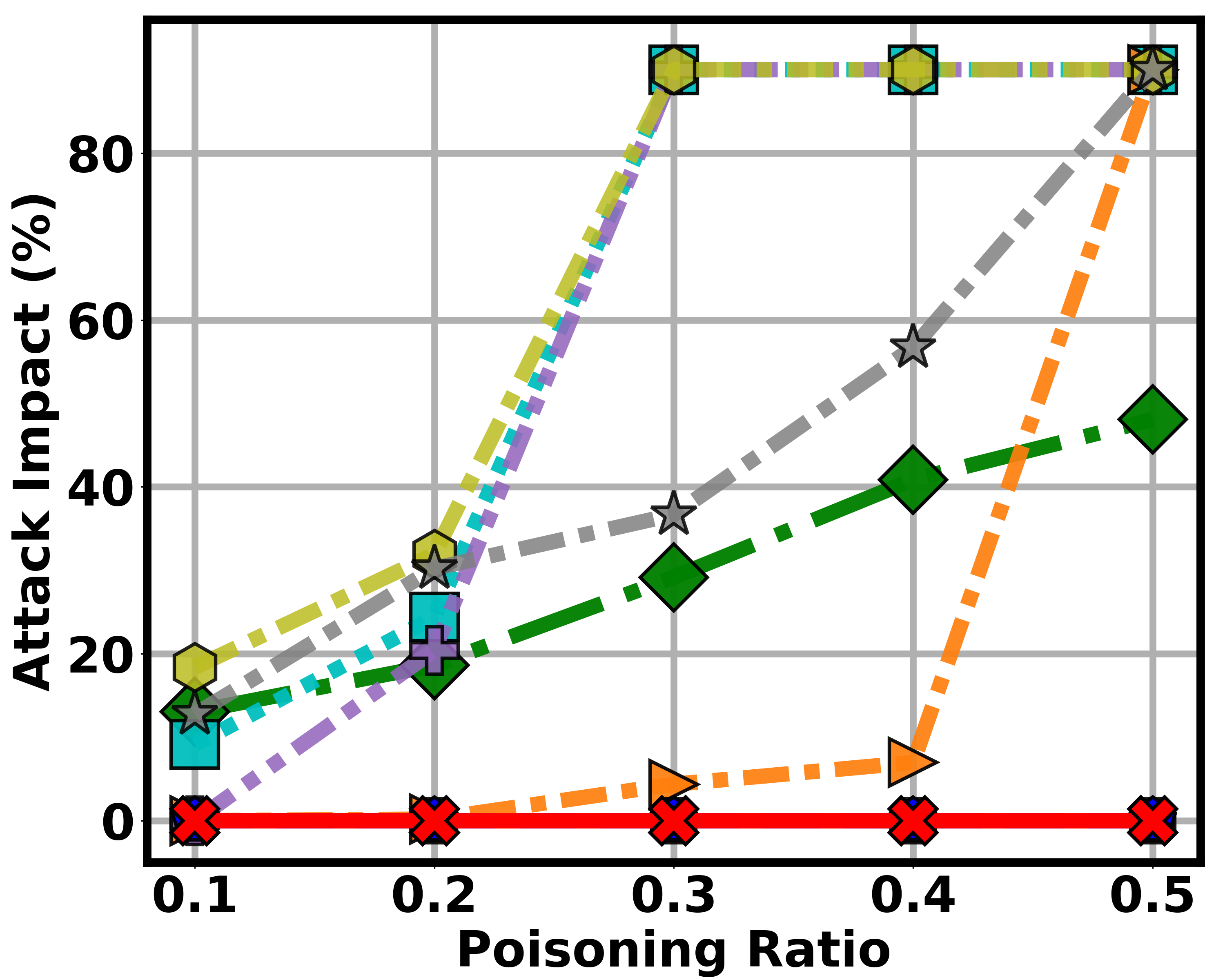}}

    \end{overpic}
    \caption{
    Comparison of defense mechanisms on three datasets under different attack strategies. 
    Although no single defense dominates across all scenarios, Hybrid-NR and Hybrid-R are robust in most cases. 
    \label{fig:attack_round}}
\end{figure*}

Next, we experimentally verify the necessity of using hybrid defense under the attack-agnostic setting.
We evaluate the defenses reviewed in Section~\ref{section:defense} under different attacks on three different tasks with non-IID data.
The first task, UCI-HAR~\cite{anguita2013public}, is a human activity recognition dataset with a skew in the feature distribution by collecting data from 30 unique subjects. 
The other two tasks are well-known image classification tasks, namely F-MNIST and CIFAR-10 with a skew of the label distribution controlled by a Dirichlet distribution~\cite{hsu2019measuring}.  
We consider $30$ clients in total participating in the FL for $100$ communication rounds.  
For reference dataset based methods, Balance and Hybrid-R, we use the same holdout dataset of size $100$.
More details of the experimental setup can be found in Appendix~\ref{sec:exp_setup}. 
In terms of the evaluation metric, we follow previous work~\cite{shejwalkar2021manipulating, shejwalkar2022back, fang2024byzantine} and define the \textit{attack impact} metric tailored to our presentation. 
We provide the formal definition below for clarity. 
For a given task, suppose that $\psi_{o}$ denotes the test accuracy of the global model without any attackers within a given time frame. 
$\psi^*$ denotes the test accuracy of the global model under a specific attack in the same time frame.  
The attack impact is given by $I := |\psi_{o} - \psi^*|$. 
Note that the maximum attack impact may vary across different attack strategies, depending on the learning task and the specific goal of the attacker.

\highlight{Robustness Under Fixed Attack Tactics.}
We first fix the attacking algorithm during the training process and evaluate the robustness of different defenses.
The results are presented in Figure~\ref{fig:attack_round}. 
It is noticed that no single defense consistently outperforms others in all scenarios.
Meanwhile, both Hybrid-NR and Hybrid-R hold a significant advantage. 
Although they may not always outperform all other defenses, hybrid defenses consistently avoid catastrophic failures or significant accuracy drops.
We highlight some key observations from the defender's perspective as follows. 

\begin{enumerate}[label=\arabic*.]
    \item As a reference dataset based method, Balance can maintain a relatively stable performance and avoid catastrophic failure. 
        For example, Balance matches or exceeds the performance of its counterparts under IPM, Min-Max, and SF attacks. 
        Meanwhile, Hybrid-R outperforms Balance under the same set of tasks. 
        This suggests that the same reference dataset can be utilized in a better way. 
    \item As an uninformed defense, CC applies momentum clipping to enhance robustness. 
        On the other hand, ROP also employs momentum to design poisoned updates. 
        It can be observed that CC completely fails under the ROP attack in three tasks. 
        This suggests that uninformed defense may not necessarily perform well in attack-agnostic settings, as the attacker may exploit the vulnerability of the specific technique used in the defense.  
    \item Defenses that are robust against untargeted attacks may still be vulnerable to targeted attacks. 
        For example, DnC is relatively robust under untargeted attacks, including IPM, Min-Max, and SF attacks on F-MNIST when the poisoning ratio is below $50\%$. 
        In contrast, DnC is still vulnerable to targeted attacks such as 3DFed and NT. Our findings challenge the prevailing belief that defenses robust against untargeted attacks are inherently effective against targeted poisoning attacks~\cite{kairouz2021advances, bhagoji2019analyzing}. 
    \item The robustness of a defense may be restricted due to the limitations of the individual modules in the defense. 
        For example, FreqFed relies on HDBSCAN clustering to separate the benign client set and the malicious participant set. 
        Specifically, HDBSCAN is sensitive to the choice of the minimal cluster size. 
        When poisoning is small, HDBSCAN tends to group poisoned updates into benign when the algorithm lowers its threshold to form a cluster. 
        Failure can be observed under ROP attack when the poisoning ratio is $0.1$. 
        This type of vulnerability due to the limitations of clustering generally exists in clustering-based defenses. 
    \item Different defenses may have different use cases. 
        More advanced defenses do not always surpass earlier ones in effectiveness. 
        For example, though SignGuard applies sign-based analysis, norm clipping, and clustering, it may not outperform Krum under SF attack across three datasets. 
        When choosing an inappropriate defense algorithm, the defender may end up with worse performance while consuming more computational and memory resources. 
    \item The reference dataset does not necessarily boost defender performance or mitigate attack impact. 
        This may be caused by an overfitting issue when the dataset is used multiple times during training.  
        For example, both Balance and CC rely on norm-based detection, whereas the former relies on the reference dataset.  
        Balance may underperform CC under the Min-Max attack on CIFAR-10, just as Hybrid-R may underperform Hybrid-NR under the same task. 
\end{enumerate}

\begin{figure}[!tb]
    \begin{overpic}[width=\linewidth, height=0.5\linewidth]{fig/4x4.pdf}

    \put(21, 48){\intab{\bf F-MNIST}}
    \put(70, 48){\intab{\bf CIFAR-10}}

    \put(47, 5){\includegraphics[width=0.53\linewidth]{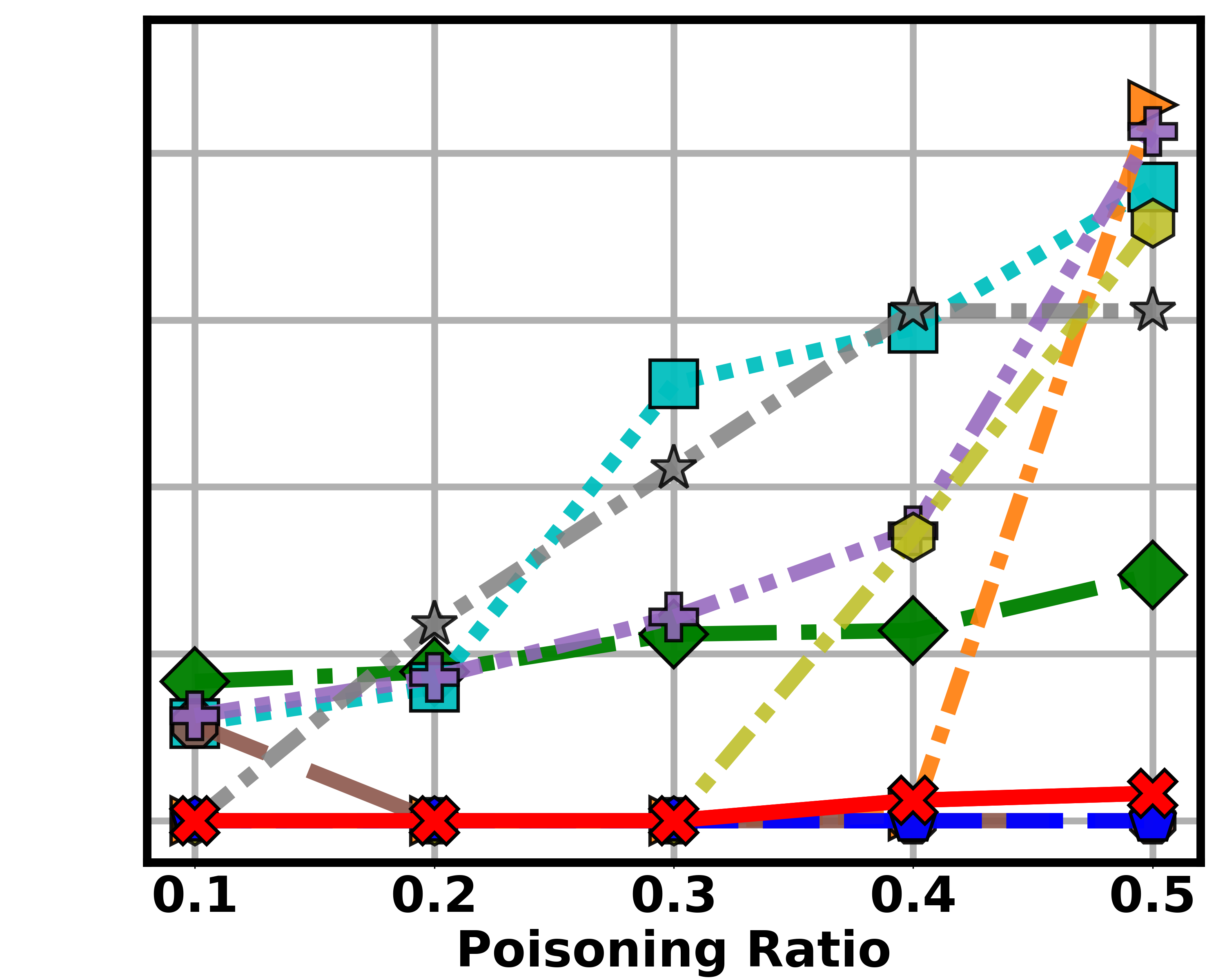}}
    \put(0, 5){\includegraphics[width=0.53\linewidth]{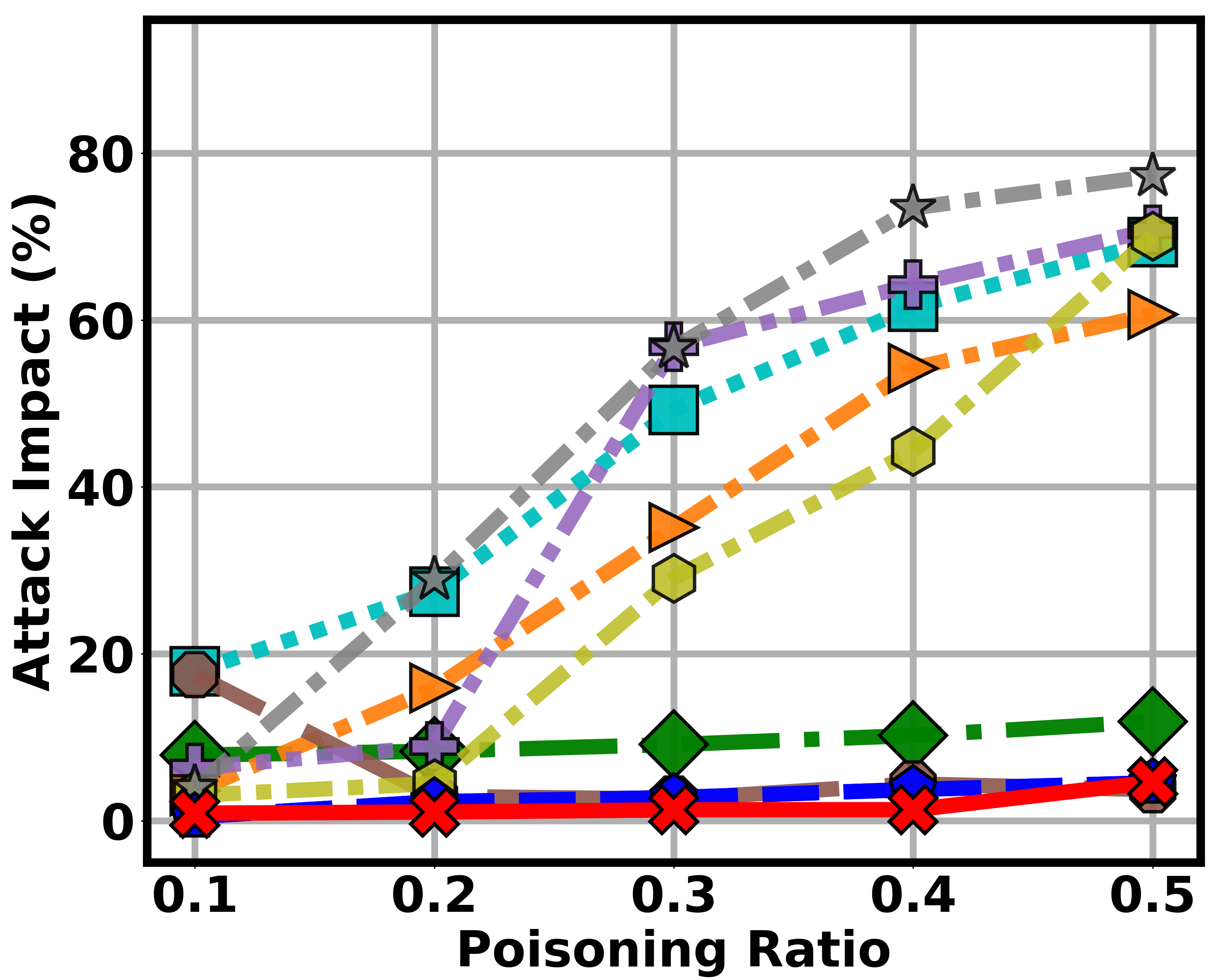}}

    \put(9, -5){\includegraphics[width=0.85\linewidth]{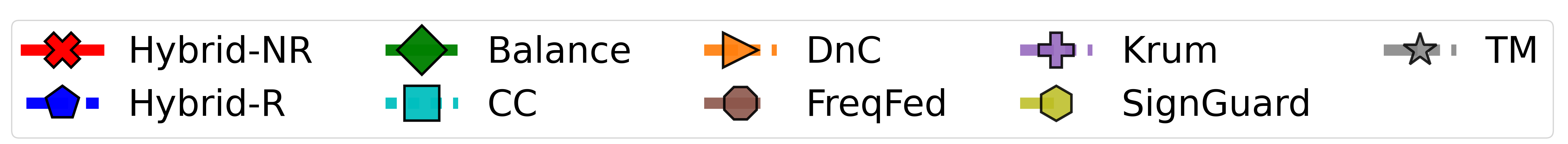}}

    \end{overpic}
    \vspace*{1pt}

    \caption{
    Mean attack impact values across different attack algorithms on the F-MNIST and CIFAR-10 tasks. The evaluation corresponds to the first attack-agnostic setting (S-1). Both Hybrid-R and Hybrid-NR rank among the most robust defenses in this scenario.  
    \label{fig:agnostic1}}
\end{figure}

\highlight{Attack-Agnostic Scenarios.}
    Building on the presentation of our hybrid defense framework, we examine three proof-of-concept scenarios within the attack-agnostic setting to highlight the necessity of the proposed defense.

(S-1) In the first scenario, we quantify the average impact across attacking tactics/algorithms. 
This setup reflects practical deployments where parallel FL training tasks may encounter region-specific adversarial behaviors. 
For example, when a multinational corporation needs to train \textbf{separate global models} for cross-regional users, different geographical regions may experience different adversarial behaviors.
In the evaluation stage, the company may seek to summarize the overall robustness of its models in a white paper or internal report. 
A high-level performance assessment across various attack strategies helps present a transparent overview that is useful to stakeholders and regulatory compliance. 
By quantifying the average impact, this scenario provides meaningful insights into the general reliability of multiple models collected from different regions.

(S-2) The second scenario considers training \textbf{one global model} under attacks launched by multiple adversarial groups, each employing different attack strategies. These groups may share a unified goal to disrupt model convergence. 
For example, one group might manipulate the signs of the gradients, while another adjusts their magnitudes.
Alternatively, these groups may pursue conflicting objectives, such as one executing an untargeted attack while another focusing on targeted manipulation. 
Such situations could arise when adversaries represent competing entities, each with motives misaligned with those of the FL organizers~\cite{ruadulescu2020multi}.
For simplicity, we use ``attack-1 + attack-2'' as an abbreviation for this scenario.

(S-3) In the third scenario, the adversaries adapt their tactics to the communication rounds. 
This dynamic behavior may result from the server sampling different participants in each round.
It can also be viewed as a temporal analogy of (S-2).
In addition, adversaries can adjust their strategies in response to evolving defense on the server side, which leads to the temporal change. 
For simplicity, we use ``attack-1/attack-2'' as an abbreviation for this scenario.

\begin{figure}[!tb]
    \begin{overpic}[width=\linewidth, height=0.5\linewidth]{fig/4x4.pdf}

    \put(21, 48){\intab{\bf NT + IPM}}
    \put(70, 48){\intab{\bf NT + ROP}}

    \put(47, 5){\includegraphics[width=0.53\linewidth]{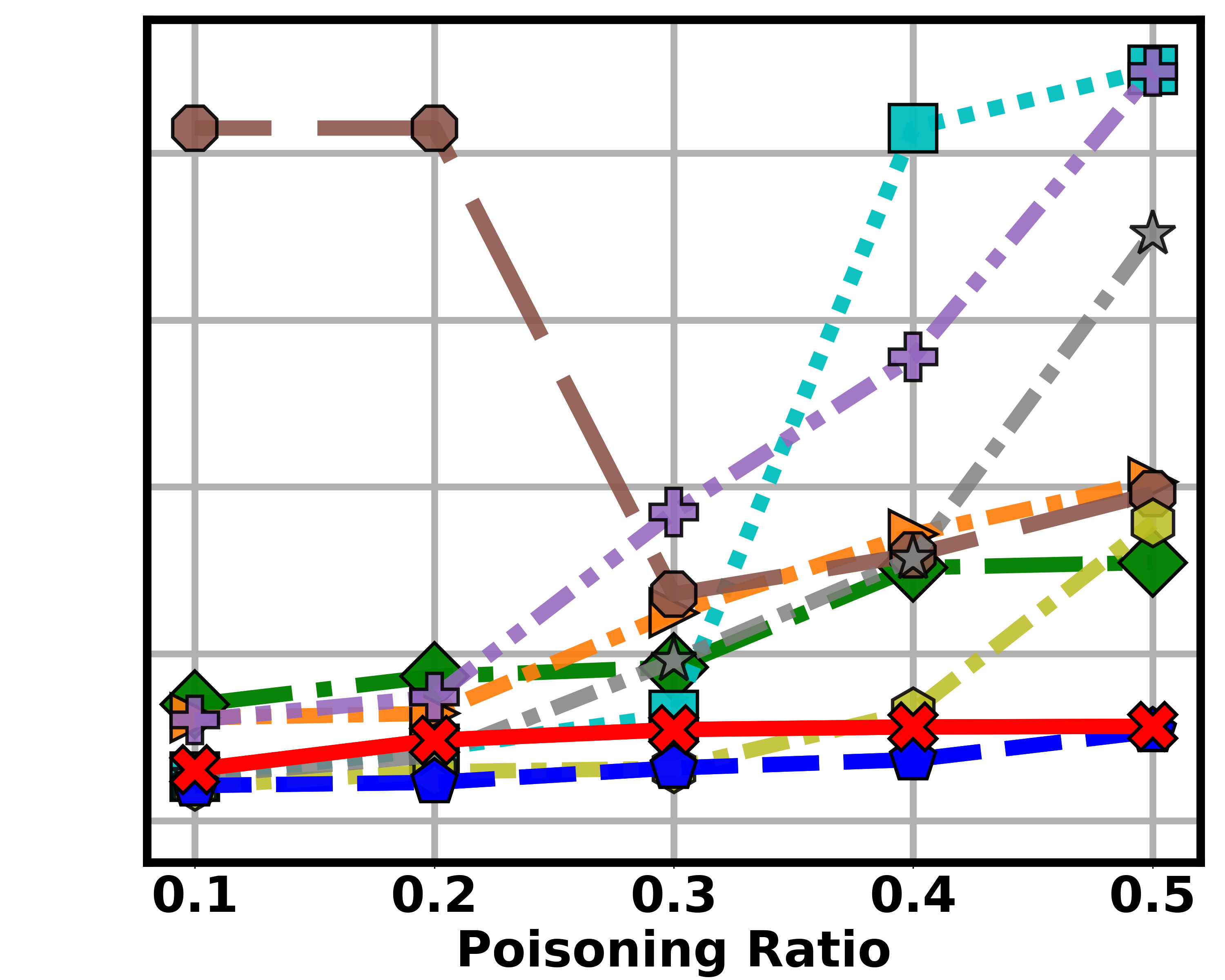}}
    \put(0, 5){\includegraphics[width=0.53\linewidth]{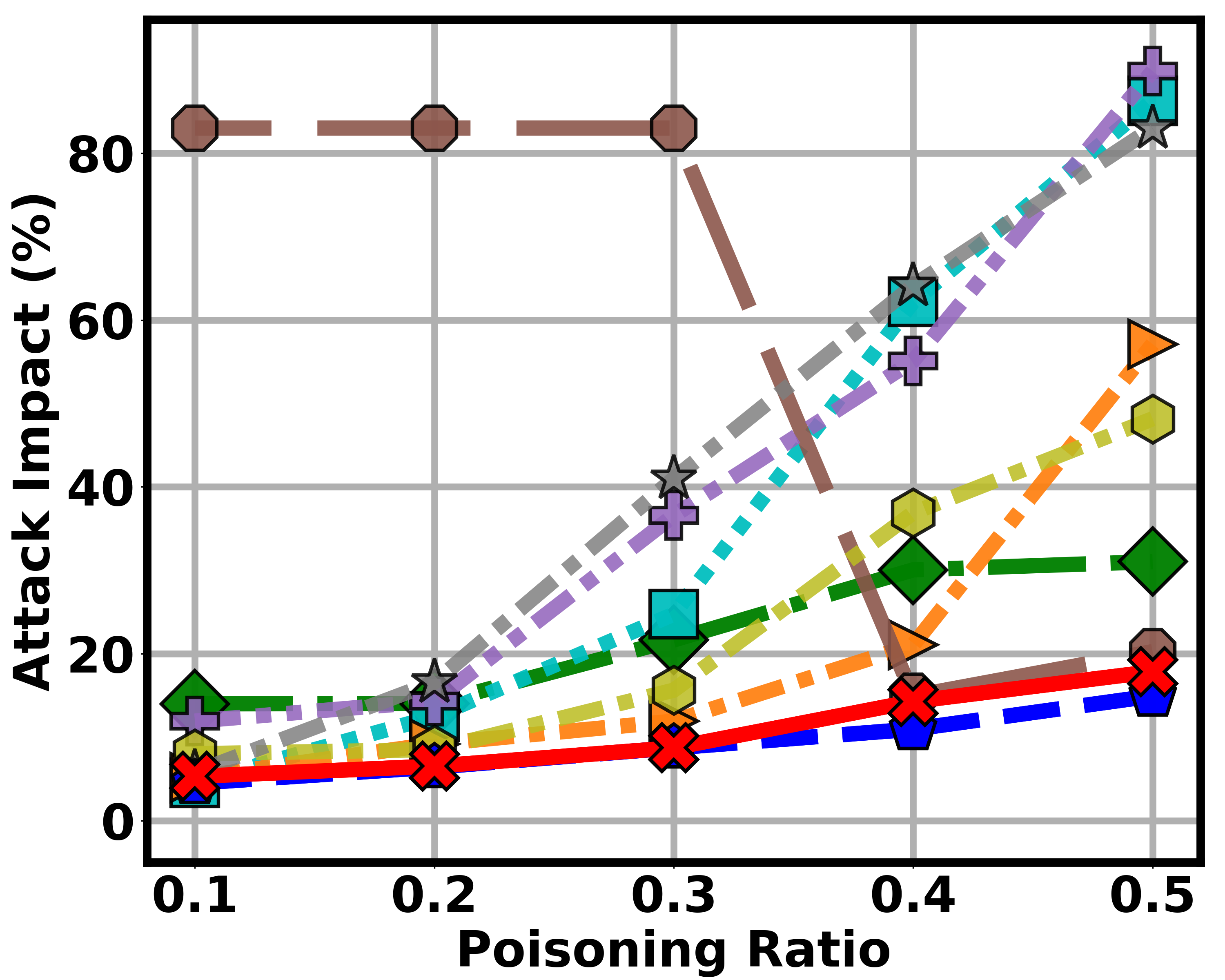}}

    \put(9, -5){\includegraphics[width=0.85\linewidth]{fig/defense_legend_2r.png}}

    \end{overpic}
    \vspace*{1pt}

    \caption{
        Comparison of defenses under the second scenario (S-2), where two groups of attackers employ different tactics. 
        Hybrid-R and Hybrid-NR demonstrate superior performance, consistently outperforming all other defenses and effectively preventing attack success.
    \label{fig:agnostic2}}
\end{figure}

Figure~\ref{fig:agnostic1} presents the scenario (S-1) with the mean impact of the attack in different attack algorithms.
Hybrid-R and Hybrid-NR consistently exhibit strong resilience in all poisoning ratios, maintaining attack impacts well below 10\% even at higher poisoning levels when $A/M \in [0.3, 0.5]$.
Meanwhile, some other defenses also show resilience in specific ranges. 
For example, FreqFed demonstrates resilience with a poisoning ratio $A/M \in [0.2, 0.5]$. 
Despite its observed robustness, FreqFed's vulnerabilities are exposed under the SF attack (Figure~\ref{fig:attack_round}). 
In parallel, Balance performs well in mitigating attack impacts below $20\%$ on the F-MNIST task. 
However, its performance degrades more in the more complicated CIFAR-10 task. 
This highlights the superior resilience of Hybrid-R and Hybrid-NR compared to other defenses.

In the second scenario (S-2), we evaluate defenses by letting two groups of attackers employ distinct attack tactics with the same communication round. 
Each group independently adopts a specific attack algorithm to achieve its objectives. 
We study two combinations of attack tactics, NT combined with IPM and NT combined with ROP. 
The poisoning attackers are evenly distributed between the two groups. 
This setup highlights the challenges of defending against adversaries that coordinate different attack strategies.
Under the ``NT + IPM'' attack, Hybrid-R and Hybrid-NR effectively mitigate the attack impact and outperform all other defenses. 
In particular, FreqFed is resistant to NT and IPM when these attacks are conducted individually (Figure~\ref{fig:attack_round}). 
However, the combination of NT and IPM causes the clustering module within FreqFed to fail, leading to significantly degraded performance.
This indicates that letting all attackers use the same poisoned update may be less effective under clustering-based methods. 

Another observation is that the combination of attacks does not always result in stronger attack impacts. 
For example, Balance demonstrates a higher attack impact (exceeding $40\%$) under a standalone NT attack when the poisoning ratio $A/M \in [0.2,0.5]$ (Figure~\ref{fig:attack_round}). 
However, when faced with combined attacks ``NT + ROP'' or ``NT + IPM'', the impact of the attack remains below $40\%$.

Note that these case studies serve as a proof of concept for the attack-agnostic settings. 
The primary goal is to demonstrate the existence of such scenarios in which existing state-of-the-art defenses fail, necessitating the adoption of hybrid defenses to achieve resilience. These results also illustrate that uninformed defenders, such as Balance and CC, may fail under practical attack-agnostic scenarios.
Other combinations of the attacks are discussed in Appendix~\ref{app:add_exp}. 
A more comprehensive analysis to include more groups, uneven attacker distributions and broader ranges of attack combinations, is left for future studies.

\begin{figure}[!tb]
    \begin{overpic}[width=\linewidth, height=0.5\linewidth]{fig/4x4.pdf}

    \put(21, 48){\intab{\bf 3DFed / IPM}}
    \put(70, 48){\intab{\bf ROP / SF}}

    \put(47, 5){\includegraphics[width=0.53\linewidth]{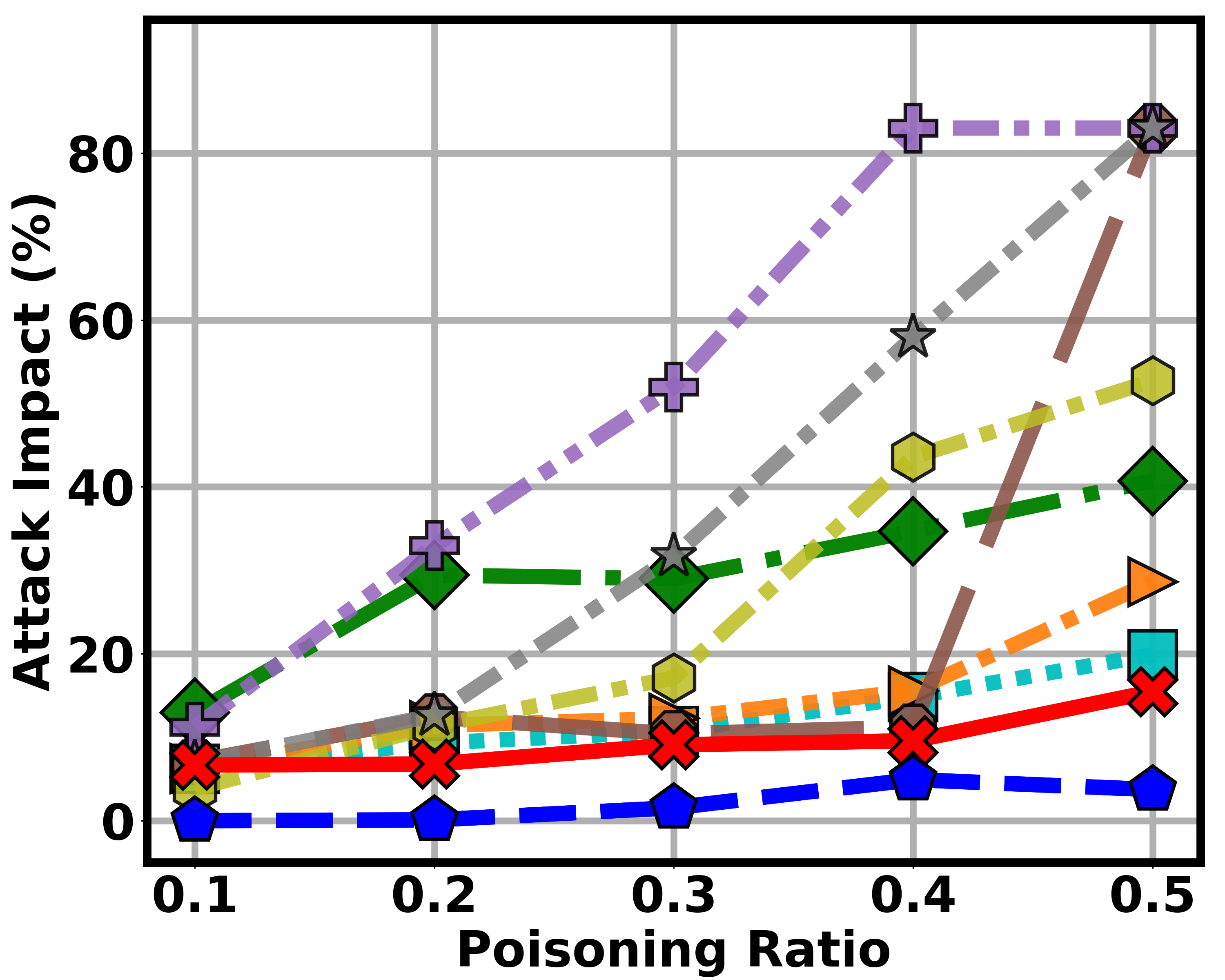}}
    \put(0, 5){\includegraphics[width=0.53\linewidth]{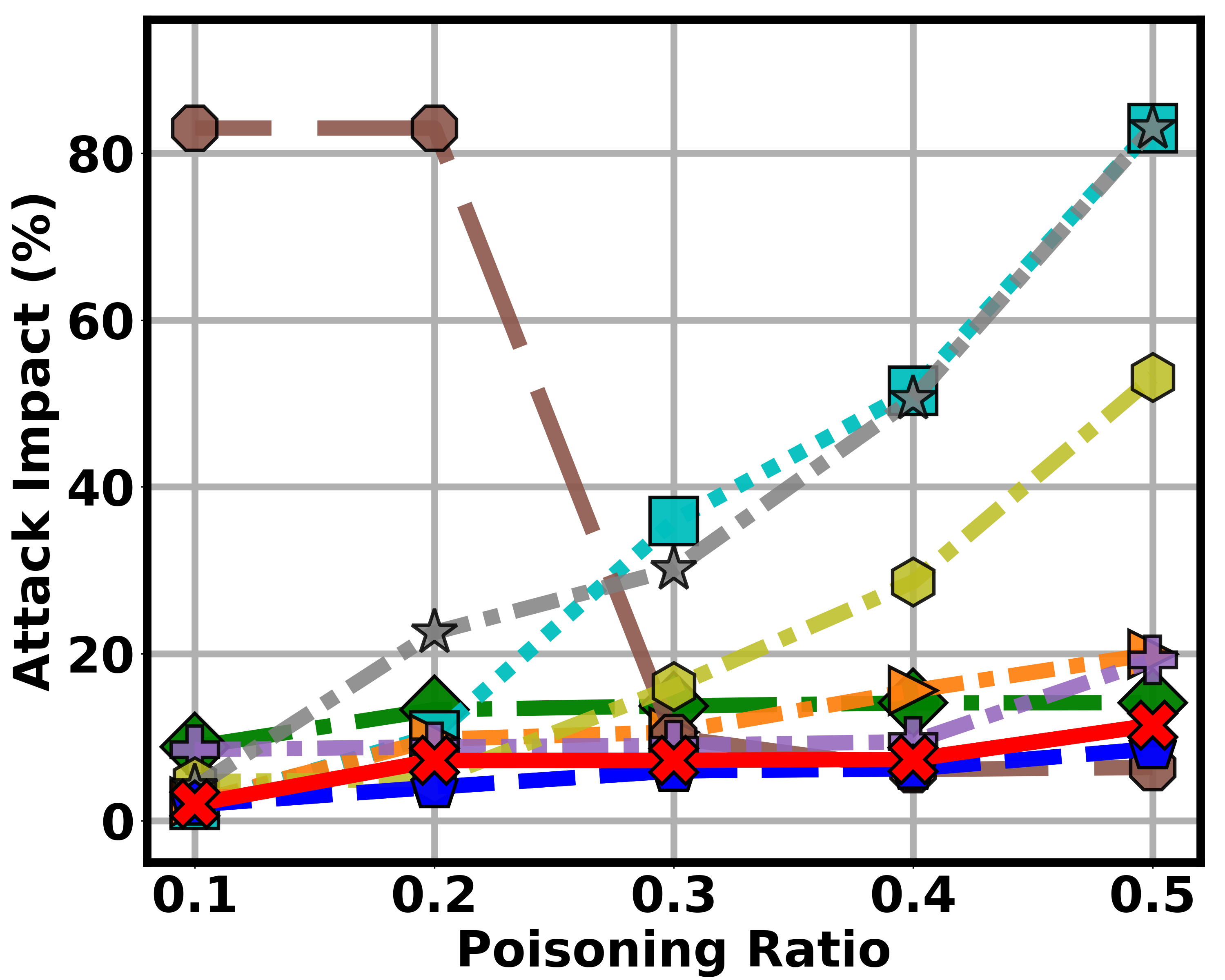}}

    \put(9, -5){\includegraphics[width=0.85\linewidth]{fig/defense_legend_2r.png}}

    \end{overpic}
    \vspace*{1pt}

    \caption{
        Comparison of defenses under the third scenario (S-3), where attack tactics switch between successive communication rounds. 
        The robustness of Hybrid-R and Hybrid-NR remains consistent with previous scenarios. Compared to other methods, the hybrid defenses effectively mitigate severe attack impacts, demonstrating their versatility as general-purpose solutions. 
    \label{fig:agnostic3}}
\end{figure}

Figure~\ref{fig:agnostic3} demonstrates the impact of the attack in the alternating attack scenarios (S-3). 
In both settings, Hybrid-R and Hybrid-NR maintain strong resilience, keeping the attack impact low even as the poisoning ratio increases. 
These hybrid defenses consistently outperform baseline methods, showcasing their ability to adapt to temporal variations in attack strategies.

\highlight{Time Complexity Analysis.}
The hybrid defense naturally increases time complexity, which can be viewed as a cost for performance improvement. 
The complexity is dominated by the most time-consuming algorithms in the defense set $\mathcal{M}$. 
Some defense mechanisms are relatively low-cost. 
For example, CC and FedAvg have a time complexity $O(d M)$, where $d$ is the dimension of model weight and $M$ is the number of clients, with $d \gg M$ in a cross-silo setting.   
Median and TrimmedMean require sorting across clients for each coordinate of the weight, which has a time complexity $O(d M \log M)$. 
In comparison, Krum has a pairwise comparison among clients, and the corresponding complexity is $O(d M^2)$. 
DnC involves eigenanalysis on a sparsified matrix,
with a time complexity of $O(dM)$.
We compare the computational time of the Krum, DnC, FreqFed, and hybrid defenses in Table~\ref{tab:time_complexity}. 
In this study, hybrid defenses are executed sequentially, but can be easily parallelized by running the algorithms in set $\mathcal{M}$ independently.

\begin{table}
\centering 
\caption{Execution Time of Various Defenses (Seconds). }\label{tab:time_complexity}
\begin{tabular}{p{0.1\linewidth}p{0.1\linewidth}p{0.1\linewidth}p{0.1\linewidth}p{0.1\linewidth}p{0.1\linewidth}}
    \toprule 
            & \intab[0.9]{Krum} & \intab[0.9]{DnC} & \intab[0.9]{FreqFed} & \intab[0.9]{Hybrid-R} & \intab[0.9]{Hybrid-NR} \\ \midrule 
    \intab[0.9]{UCI-HAR} & $ 0.14 $ & $0.08 $ & $0.12$ & $ 0.48 $ &  $0.55$ \\ \cmidrule{1-6} 
    \intab[0.9]{F-MNIST} & $ 0.16 $ & $0.22 $ & $0.13$ & $0.75 $ & $0.83$ \\  \cmidrule{1-6} 
    \intab[0.9]{CIFAR-10} & $  0.30 $ & $0.41$ & $0.25$ & $1.33 $ & $1.57$ \\ \bottomrule
\end{tabular}
\end{table}

\highlight{Merits of Hybrid Defenses.}
We summarize the merits of the hybrid defender as follows. 
By uniting the features of the existing robust aggregation strategies we reviewed in Section~\ref{section:defense}, the defender is able to 
\begin{enumerate}[itemindent=2em]
    \item[(D-1)] Remove poisoned gradients that are far from benign gradients or neighbors. 
    This may be achieved by adopting robust statistics or by calculating outlier indicators. 
    \item[(D-2)] Reject the gradients that point directly in the opposite direction to the benign gradients.  
    This may be achieved by analyzing sign statistics or score related to the inner product.  
    \item[(D-3)] Mitigate the impact of subtle noise introduced by adversaries.
    Using frequency domain filtering, the defender has improved resilience against noise disturbance. 
\end{enumerate}

\section{TrapSetter Attack}\label{section:attack}

In this section, we examine the vulnerabilities of hybrid defenses from the attacker's perspective. 
To this end, we propose a novel attack algorithm, \textit{TrapSetter}, designed to explore the limitations of hybrid defenses. 
As described in Section~\ref{sec:hybrid_defense}, hybrid defenses demonstrate resilience against norm-based poisoning, sign manipulation, and noise injection. 
Although these defenses represent state-of-the-art protection in FL, they are not a panacea against all adversarial threats.
In this section, we introduce the TrapSetter attack, which serves as a reminder of constant vigilance in the everlasting arms race between the attacker and the defender.

The proposed TrapSetter attack is designed as follows, with the pseudocode summarized in Algorithm~\ref{alg:trapsetter}.
Following the assumptions of the threat model in Section~\ref{section:preliminaries}, we assume that the adversaries have access to a dataset $\mathcal{D}_{\mathcal{A}}$. 
The dataset $\mathcal{D}_{\mathcal{A}}$ may be split into two parts, a training set $\mathcal{D}_{\mathcal{A}}^{\text{train}}$ and a validation set $\mathcal{D}_{\mathcal{A}}^{\text{val}}$, for different purposes. 
An attacker initially estimates a global weight in  $\tw[k+1]_{a}$ by using a standard aggregation rule, such as FedAvg, 
\begin{equation}
    \tw[k+1]_{a} \triangleq \w[k] - \frac{\eta}{n} \sum_{t=0}^{\tau-1}\! \nabla R (\w[k,t]_{a}; \xi^{(k,t)}_{a} ), 
\end{equation}
where $a$ is the attacker index, $ \xi^{(k,t)}_{a}$ is the local batch with size $n$ sampled from $\mathcal{D}_{\mathcal{A}}^{\text{train}}$.
Using $\tw[k+1]_{a}$ as a reference, the attacker looks for a trap weight $\hw[k]_{a}$ that is close to $\tw[k+1]_{a}$ but leads to the worst performance evaluated on the attackers' dataset. 
Specifically, the attacker selects two directions, $\vp_{1}^{(k)}$ and $\vp_{2}^{(k)}$, and perturb the weight $\tw[k+1]_{a}$ along the two directions and perform grid searching. 
For simplicity, the index $a$ is omitted for the perturbation vectors. 
In practice, multiple attackers may use distinguished vectors to reduce the chance of being detected by clustering-based defense.  
The concept of selecting two direction vectors to explore the objective function space has been extensively explored in the literature~\cite{li2018visualizing,ozfatura2023byzantines}. 
This approach is considered a balanced design, offering a trade-off between exploration flexibility and computational complexity. 

\begin{algorithm}[!tb]
  \SetKw{init}{Initialize}
  \SetKw{inputs}{Inputs}

  \SetKwBlock{trapsetter}{Function Attack $(\w, r)$}{return $\w_{\text{trap}}$}
  \SetKwBlock{trap}{Function Trap$(\tw[k+1], r, \mathcal{D})$}{return $\w_{\textrm{trap}}$}

  \caption{TrapSetter Attack \label{alg:trapsetter}}

  \KwIn{\textrm{Attackers' reference dataset} $\mathcal{D}_{\mathcal{A}} = \mathcal{D}_{\mathcal{A}}^{\text{train}} \cup \mathcal{D}_{\mathcal{A}}^{\text{val}}$, 
  global weight $\w[k]$} 

  \For{ \textrm{\normalfont  each attacker} $a \in \mathcal{A}$ }{  
      Initialize $\w[k, 0]_{a} \leftarrow \w[k]$  \\
      \For{$t \in \{0, 1, \dots, \tau-1\}$}{
          $\xi^{(k,t)}_{m} \sim \mathcal{D}_{\mathcal{A}}^{\text{train}} $  \hfill $\triangleright$ \text{sample a mini-batch} \\
          $\w[k, t+1]_{a} = \w[k, t]_{a} - \frac{\eta}{n_{a}} \nabla R\left(\mathbf{w}_{a}^{(k, t)} ; \xi_{a}^{(k, t)}\right)$ \\ 
      }
      $\displaystyle \tw[k+1]_{a} \leftarrow \w[k, \tau]_{a} $  \\
      sample $\zeta_{a} \sim \mathcal{U}(\zeta_{L}, \zeta_{U}) $, $r_{a} \sim \mathcal{U}(r_{L}, r_{U})$ \\
      $\hw[k]_{a} \leftarrow  \text { Trap}\left(\tw[k+1]_{a}, \, r_{a}, \, \mathcal{D}_{\mathcal{A}}^{\text{val}} \right)$  \\
      $\displaystyle \update[k]_a \!=\! \frac{\zeta_{a}}{A} \! \left[ M  \w[k] \!-\! M \hw[k]_{a} \!-\! \frac{B}{\zeta_{a}} (\w[k] \!-\! \tw[k+1]_{a} ) \right]$
  } 
  \Return{$\update[k]_a$}

  \vspace*{5pt}
  \trap{
      \init $\vp^{(k)}_{1}$, $\vp^{(k)}_{2}$, step size $\delta_{r}$, matrix $\boldsymbol{\Psi}$ \\
      \For {$\kappa^{(i)}_{1}, \kappa^{(j)}_{2} \in \{-r, -r + \delta_{r}, \dots, r- \delta_{r}, r \}$}{
      $\displaystyle \hw[k] = \tw[k+1] +  \kappa^{(i)}_1 \vp_{1}^{(k)} + \kappa^{(j)}_2 \vp_{2}^{(k)}$ \\
      \textbf{if} $\|\hw[k] - \tw[k+1] \| > r$: \textbf{continue} \\
      $\Psi_{i,j} \leftarrow \psi(\hw[k], \mathcal{D})$ \\ 
      }
      $i^*, j^* = \argmin_{i,j} \Psi_{i,j}$ \\
      $\hw[k] = \tw[k+1] +  \kappa^{(i^*)}_1 \vp_{1}+ \kappa^{(j^*)}_2 \vp_{2}^{(k)}$ \\
      \Return{$\hw[k]$}
  }
\end{algorithm}

The first direction vector  $\vp_{1}^{(k)}$ is set as a normalized vector opposite to aggregated benign gradients,
\begin{equation}
    \vp_{1}^{(k)} = -\frac{ \w[k] - \tw[k+1]_{a} }{\left\|  \w[k] - \tw[k+1]_{a} \right\|_2 }.
\end{equation}
The second direction vector $\vp_{2}^{(k)}$ is a normalized Gaussian noise vector,
\begin{equation}
    \vp_{2}^{(k)} = \frac{\mathbf{n}}{\left\|\mathbf{n} \right\|_2}, \quad \mathbf{n} \sim \mathcal{N}(\mathbf{0}, \varsigma^2 \mathbf{I}),
\end{equation}
where $\varsigma$ is a positive constant. 
A trap weight $\hw[k]$ may be constructed as 
\begin{equation}
    \hw[k]_a = \tw[k+1]_{a} +  \kappa_1 \vp_{1}^{(k)} + \kappa_2 \vp_{2}^{(k)}, 
\end{equation}
where $\kappa_{1}, \kappa_{2} \in [-r, r]$ are scaling factors, with $r$ denoted as a radius parameter that controls the distance between $\hw[k]_{a}$ and $\tw[k+1]_{a}$.
The trap weight can be found by solving the following optimization problem:
\begin{subequations}\label{eq:r_restrcition}
\begin{align}
    & \min_{\kappa_1, \kappa_2 } \; \psi\left(\hw[k]_{a} ; \mathcal{D}_{\mathcal{A}}^{\text{val}} \right), \\
    & \text{s.t. } \| \hw[k]_{a} - \tw[k+1]_{a} \|_2 \leqslant r,  
\end{align}
\end{subequations}
where $\psi(\w; \mathcal{D})$ is an accuracy function evaluating a model $\w$ on a dataset $\mathcal{D}$. 
One can approximate a solution to the problem by dividing the feasible region $[-r, r] \times [-r, r]$ into a mesh grid and assessing the accuracy at each grid point to identify an optimal  $\hw[k]_{a}$.
The optimization procedure is detailed in Algorithm~\ref{alg:trapsetter} starting from Line 9.
After obtaining the trap weight $\hw[k]_{a}$, the poisoned update $\update[k]_a$ is then crafted as follows: 
\begin{equation}\label{eq:trapsetter_update}
    \update[k]_a \!=\! \frac{\zeta}{A} \! \left[ M  \w[k] \!-\! M \hw[k]_{a} \!-\! \frac{B}{\zeta} (\w[k] \!-\! \tw[k+1]_{a} ) \right], 
\end{equation}
where $\zeta $ is a positive scaling factor. 
We now look into the intuition of designing the poisoned updates. 
We first assume that the attacker gives an unbiased estimation of benign gradients, 
\begin{equation}
    \mathbb{E}[\w[k] - \tw[k+1]_{a} ] = \mathbb{E}[\w[k] - \w[k+1]_{b}] = \mathbb{E}[\update_{b}], 
\end{equation}
where $b \in \mathcal{B}$ is the benign client index. 
By letting $\zeta=1$, the attackers can thus mislead the server under FedAvg, 
\begin{equation}
   \mathbb{E}[\w[k+1]] = \mathbb{E}[\hw[k]_{a}].
\end{equation}
To deceive different defense algorithms, attackers can change the scaling factor $\zeta$ and the optimization radius $r$. 
For example, for norm-based defense, smaller $\zeta$ and $r$ may be more desirable. 
In our design, we let individual attackers sample the radius and scaling factors from uniform distributions.

\begin{figure}[!tb]
    \begin{overpic}[width=\linewidth, height=1.4\linewidth]{fig/4x4.pdf}
    
    \put(10, -1){\includegraphics[width=0.81\linewidth]{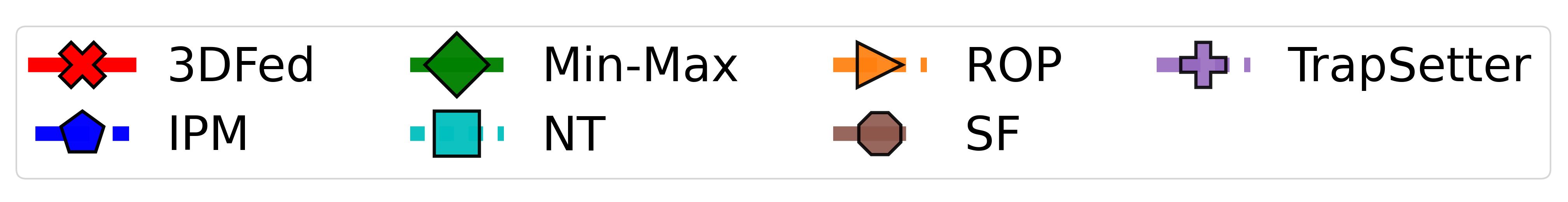}}
    
    \put(0, 78){\rotatebox{90}{\intab{\bf UCI-HAR}}}
    \put(0, 48){\rotatebox{90}{\intab{\bf F-MNIST}}}
    \put(0, 18){\rotatebox{90}{\intab{\bf CIFAR-10}}}

    \put(17, 97.5){\intab{\bf Hybrid-NR }}
    \put(50, 97.5){\intab{\bf Hybrid-R }}

    \put(35, 7){\includegraphics[width=0.517\linewidth]{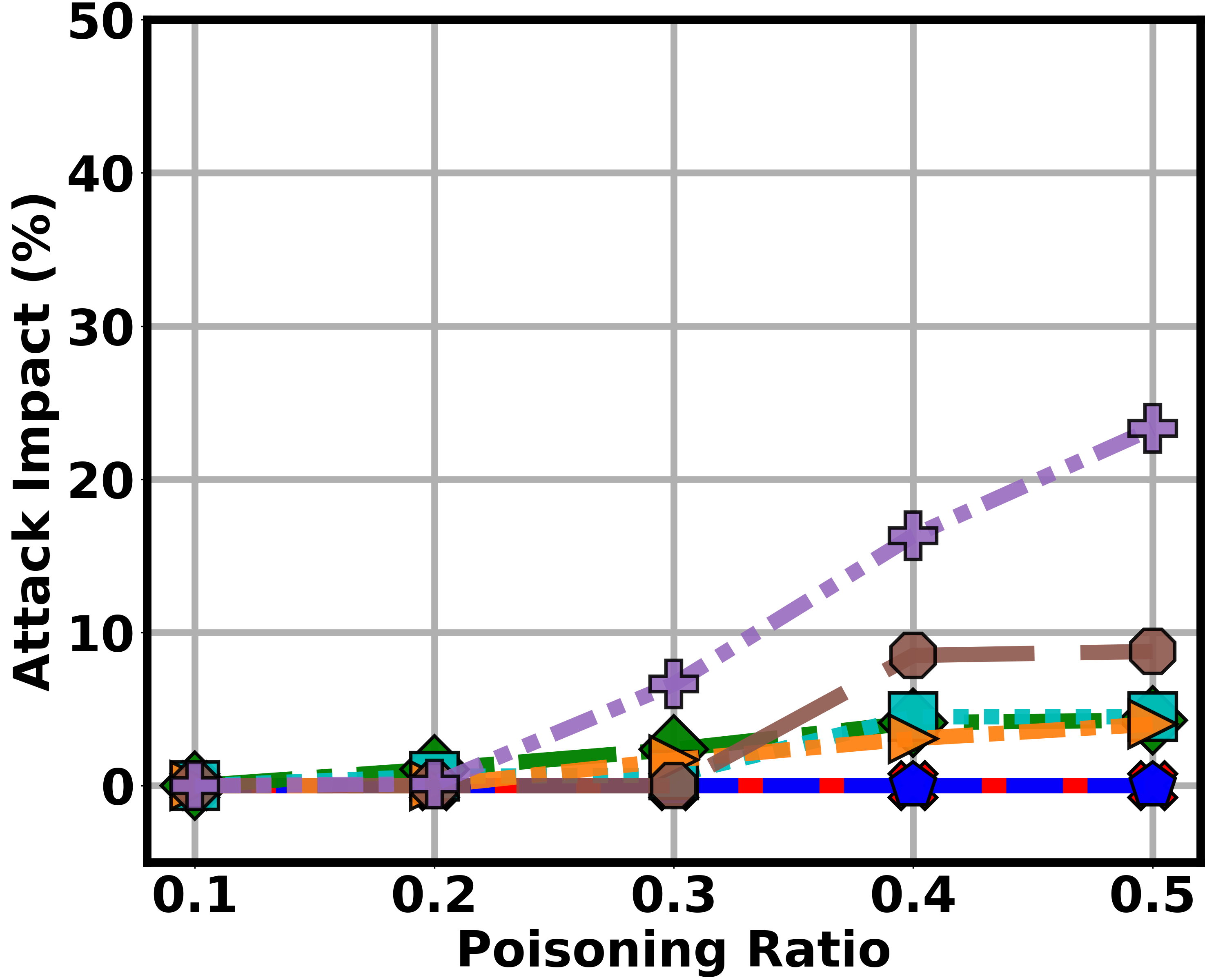}}
    \put(2, 7){\includegraphics[width=0.517\linewidth]{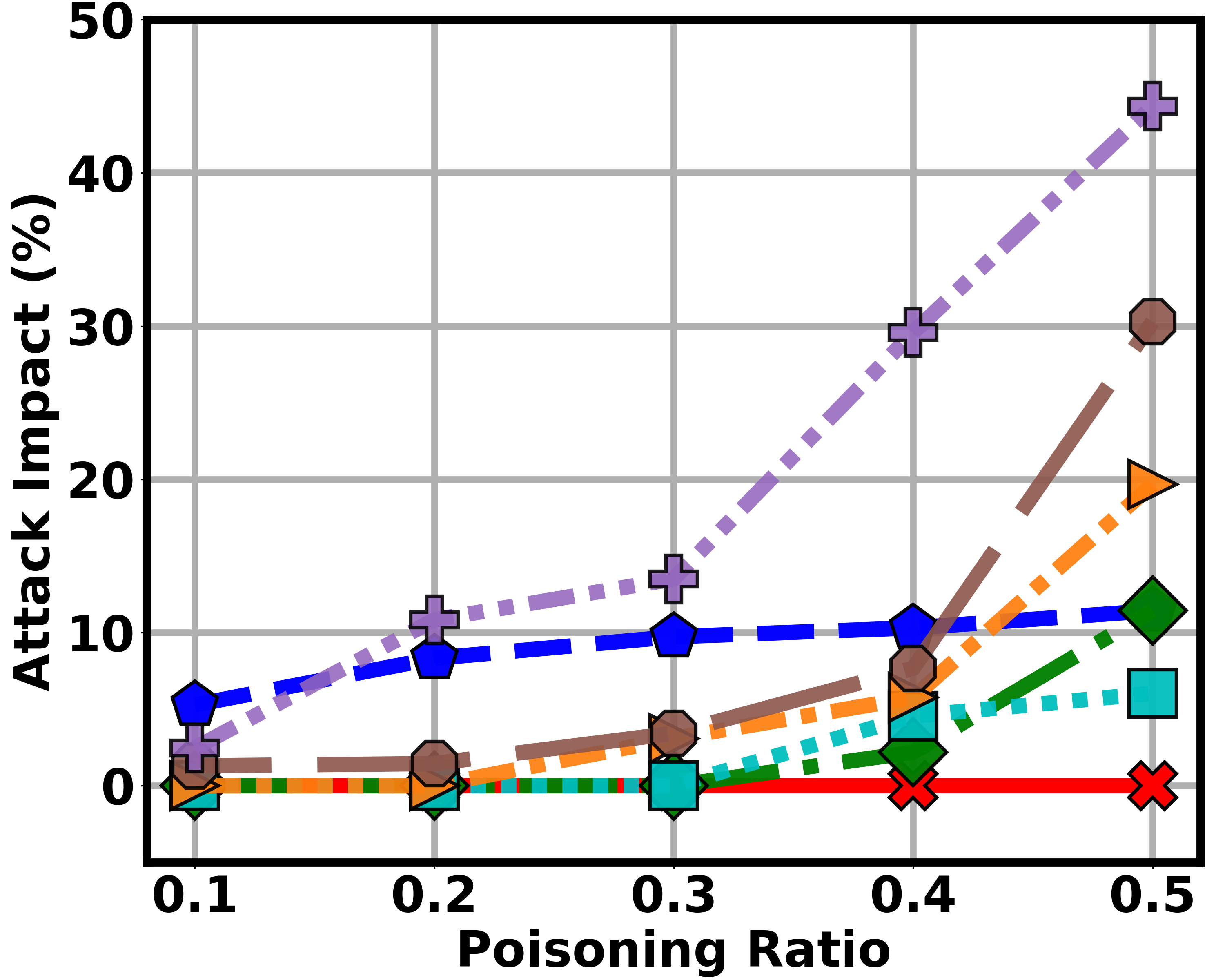}}

    \put(35, 37){\includegraphics[width=0.517\linewidth]{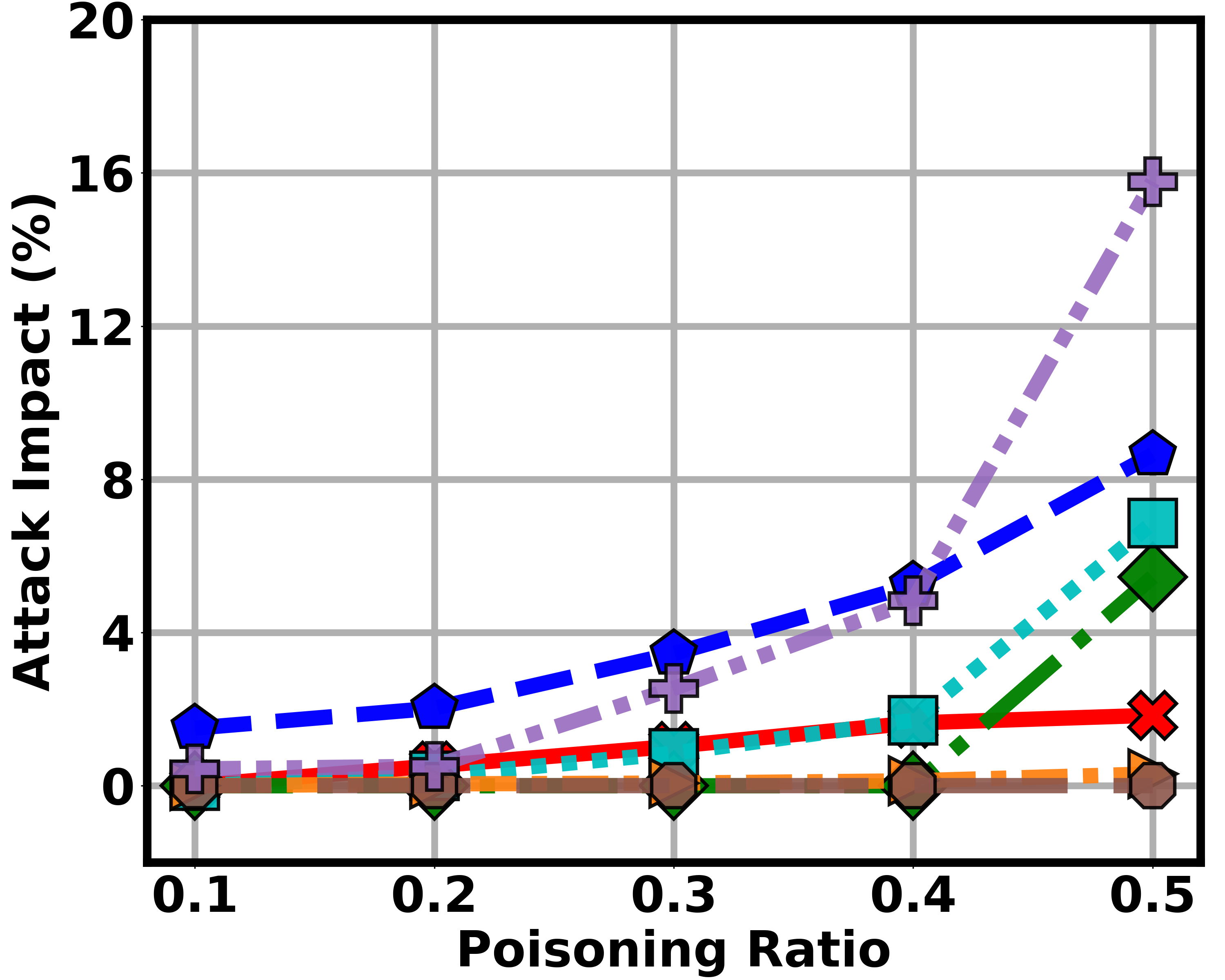}}
    \put(2, 37){\includegraphics[width=0.517\linewidth]{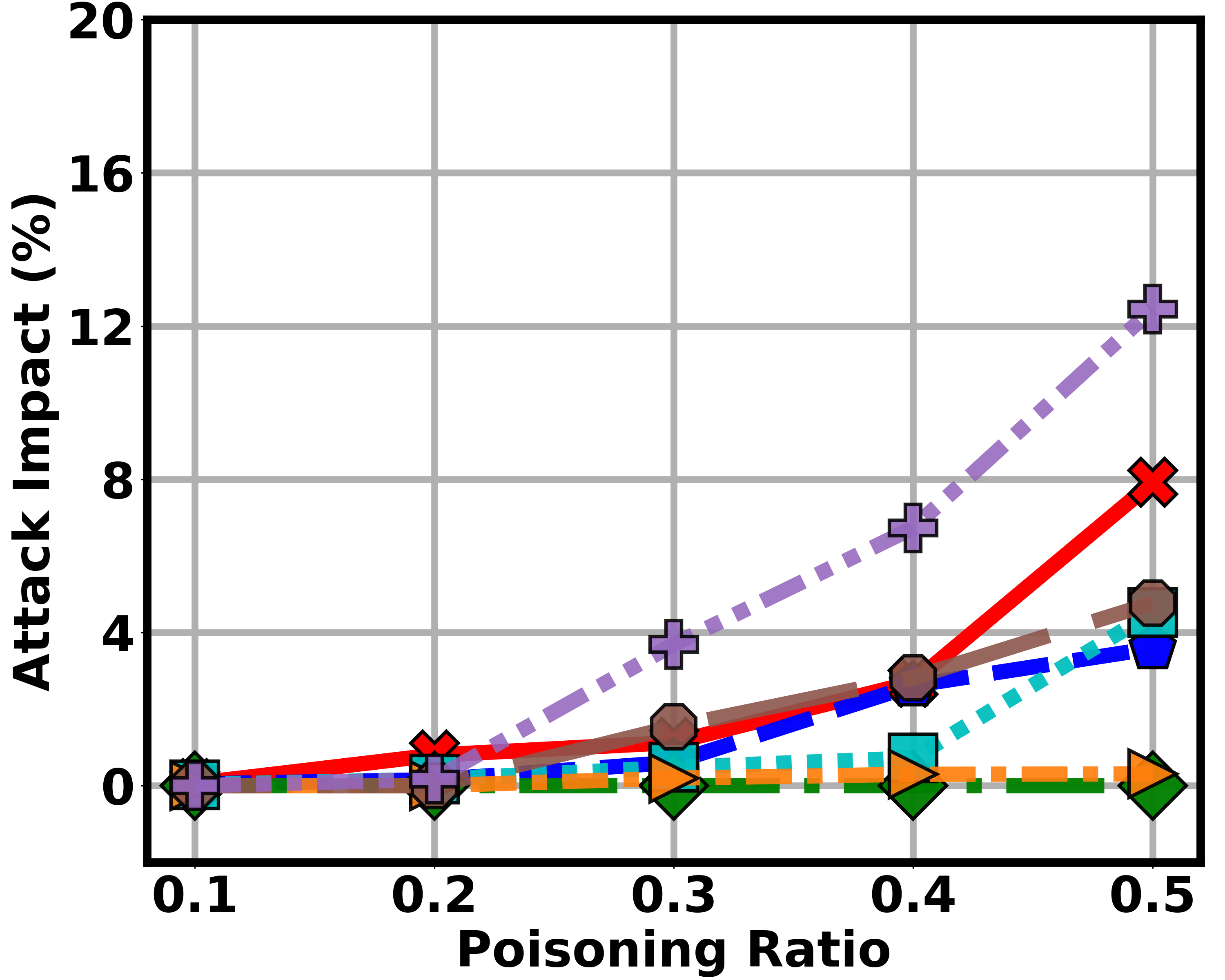}}

    \put(35, 67){\includegraphics[width=0.517\linewidth]{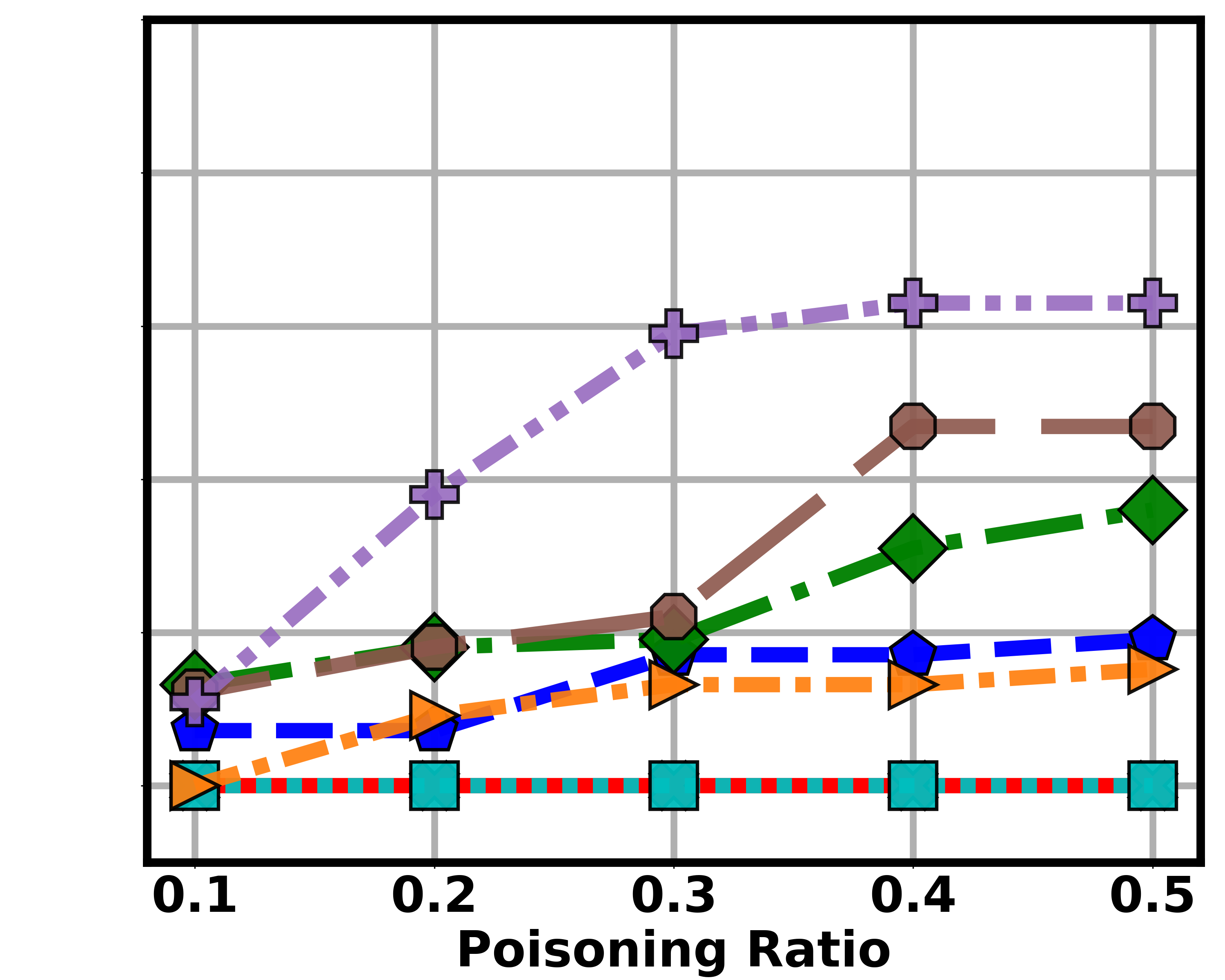}}
    \put(2, 67){\includegraphics[width=0.517\linewidth]{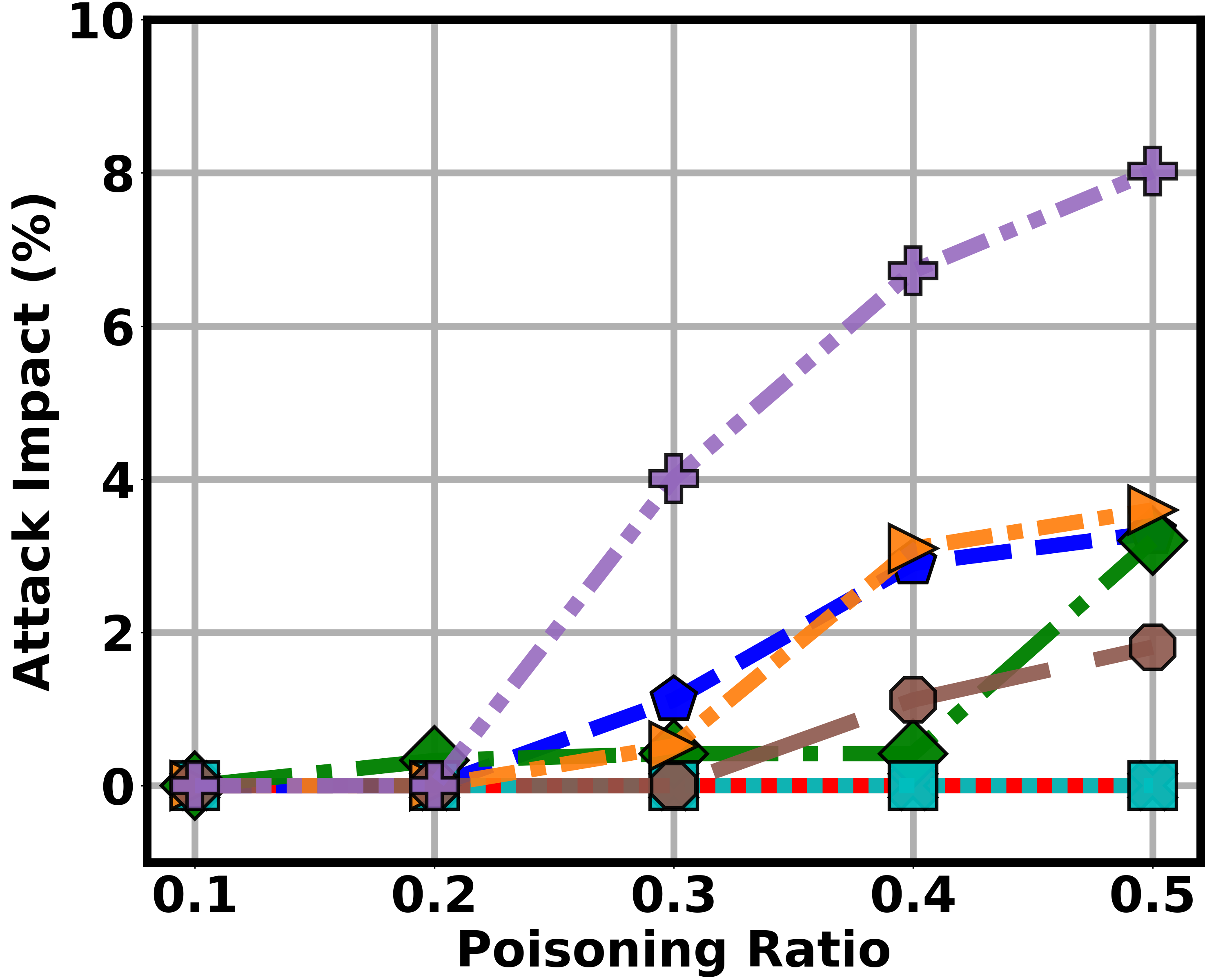}}
        
    \end{overpic}
    \caption{
    Test accuracy versus the poisoning ratio for different attacks under Hybrid-NR defense and Hybrid-R defenses. 
    Compared to other adversaries, the proposed TrapSetter attack may further increase the impact by $5\%$ to $15\%$ when the poisoning ratio is greater than $0.3$. 
    \label{fig:trap_under_hybrid}}
\end{figure}

In Figure~\ref{fig:trap_under_hybrid}, we illustrate the attack impact of various attack algorithms, including the proposed TrapSetter, under the Hybrid-NR and Hybrid-R defenses. 
TrapSetter consistently outperforms other attacks on the UCI-HAR and F-MNIST datasets. 
Under Hybrid-R defense, TrapSetter can further increase the absolute impact of attack by $5\%$ to $15\%$ compared to other state-of-the-art attacks when the poisoning ratio $A/M \in [0.3, 0.5]$,
while the impact may be slightly larger under Hybrid-NR defense. 

The impact of the attack also varies between different tasks, and the complexity of the task plays a critical role. 
For example, CIFAR-10 is generally considered more complex than F-MNIST and UCI-HAR. 
As a result, both TrapSetter and its competitors achieve potentially higher attack impacts on CIFAR-10. 
In particular, TrapSetter exhibits a consistent trend of monotonic increase in attack impact with respect to task complexity, a pattern that is not observed in other attacks. 
This suggests that the interplay between hybrid defenses and the TrapSetter attack could offer valuable insight into the relationship between the complexity of the learning task and the robustness of the model.

In this work, hybrid defenses are proposed as general-purpose solutions to mitigate Byzantine attacks. 
Meanwhile, the designed TrapSetter attack aims to boost the attack impact under hybrid defenses and investigate their vulnerabilities, serving as a reminder that constant vigilance is required. 
The TrapSetter attack can be viewed as a ``dedicated'' attack algorithm toward hybrid defenses and may not always outperform all other attacks under other defenses.  
This is due to the design in which TrapSetter explores a local minimum during its optimization process, which may limit its effectiveness against some mechanisms. 
For example, IPM may demonstrate greater efficacy under robust statistic-based defenses, such as Krum and Trimmed Mean compared to TrapSetter. 
Additional discussions and analyses are provided in the Appendix~\ref{app:add_exp}. 

\begin{figure}[!tb]
    \begin{overpic}[width=\linewidth, height=0.5\linewidth]{fig/4x4.pdf}

    \put(21, 48){\intab{\bf Balance}}
    \put(70, 48){\intab{\bf Hybrid-R}}
    
    \put(48, 5){\includegraphics[width=0.517\linewidth]{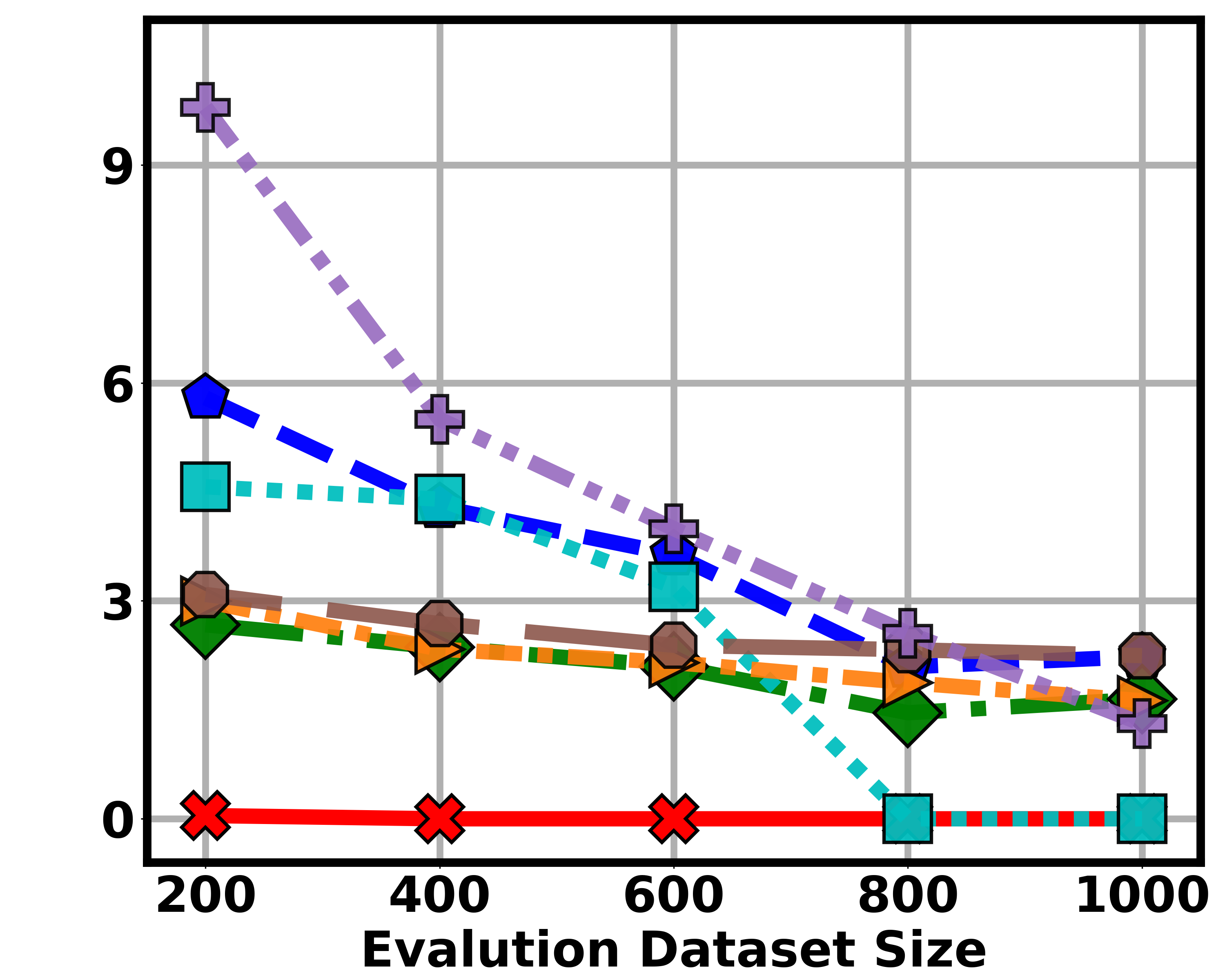}}
    \put(0, 5){\includegraphics[width=0.517\linewidth]{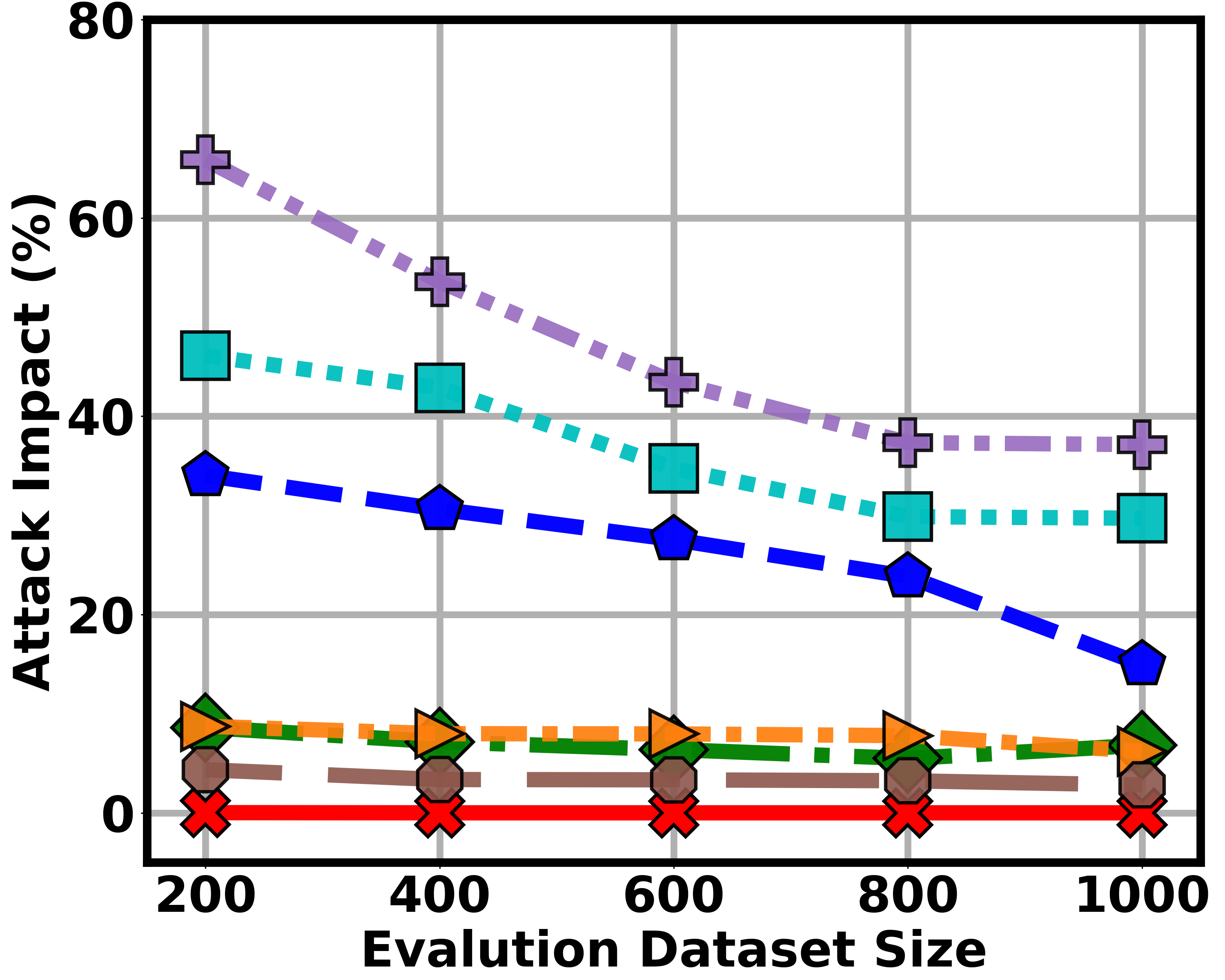}}

    \put(9, -6){\includegraphics[width=0.85\linewidth]{fig/attack/defense_legend.png}}

    \end{overpic}
    \vspace*{1pt}

    \caption{
    Attack impact versus the evaluation dataset size $|\mathcal{D}_0|$ on the F-MNIST task. 
    The impact generally decreases with the dataset size. 
    Meanwhile, the impact rankings of different attacks tend to be consistent. 
    \label{fig:root_size}}
\end{figure}

To further examine the resilience of the system under the TrapSetter attack, we shift our focus to reference dataset-based defenses. 
As a case study, we evaluated the attack impact of various attacks under two defenses based on the reference dataset, Balance and Hybrid-R. 
Fixing the poisoning ratio at $0.4$, we vary the size of the reference dataset $|\mathcal{D}_0|$. 
The results in Figure~\ref{fig:root_size} show that the attack impact decreases as the size of the reference dataset increases. In particular, the ranking of attack impacts remains consistent across the two defenses, highlighting the stability of their relative effectiveness.

\highlight{Summary of Attack Experiments.} 
We summarize three insights into the success of the TrapSetter attack as follows.
First, we have pointed out that hybrid defenses are sensitive to the magnitude deviation of poisoned gradients in (D-1). 
By adaptively changing the radius $r$ in \eqref{eq:r_restrcition} and the scaling factor $\zeta$ in~\eqref{eq:trapsetter_update}, the adversaries are able to bypass this type of detection while still making the attack impactful. 
In addition, the hybrid defense is immune to the sign manipulations in (D-2) and the injection of noise in (D-3).
Our optimization in Algorithm~\ref{alg:trapsetter} searches for a \textit{trap weight} without explicitly considering signs or injecting noise. 
This forces the defender to switch between different aggregation strategies, which can lead to a suboptimal solution.

\section{Conclusion}\label{section:conclusion}

In this paper, we have reviewed advances in poisoning attacks and defenses in FL.
We have introduced and evaluated two variants of hybrid defenses. 
The first defense, Hybrid-R, dynamically adapts its strategies using a small validation dataset. 
The second defense, Hybrid-NR, achieves robustness via dual robust aggregation.
Through extensive simulations, we demonstrated that both variants outperform existing aggregation schemes under different types of attacks.

To further validate the robustness of these defenses, we explored different attack-agnostic scenarios.
Specifically, we allow the attackers to form into different groups and use different algorithms. 
Our results highlight that as adversaries adopt more dynamic and adaptive tactics, hybrid defenses become essential to maintaining robust FL performance. 

Despite the strengths of hybrid defenses, our proposed TrapSetter attack revealed potential vulnerabilities, highlighting that robust defenses need to continuously evolve to address emerging and sophisticated threats.
TrapSetter attack successfully boosts the impact under hybrid defenses, further reducing model test accuracy by 5--15\% across diverse tasks.
We encourage the research community to continue to develop innovative defense strategies to address these evolving threats.


\bibliographystyle{IEEEtran}
\bibliography{ref}

\end{document}